\documentclass[12pt]{myarticle}
\usepackage{mycite}
\renewcommand\refname{}		

\usepackage[utf8]{inputenc}             
\usepackage{graphicx}
\usepackage{color}
\usepackage{url}                 
\usepackage{setspace}   
\definecolor{darkblue}{rgb}{0,0,0.7}
\newcommand{\cit}[1]{$^{\citen{#1}}$}
\newcommand{\dg}{\ensuremath{^\circ }}
\newcommand{\etal}{\emph{et al.}}

\newcommand{\cp}{\color{black}\ }

\newcommand\hl{\rule{\textwidth}{0.25mm}} 

\newcommand\go{\mathrel{\raise.3ex\hbox{$>$}\mkern-14mu 
             \lower0.6ex\hbox{$\sim$}}}
\newcommand\lo{\mathrel{\raise.3ex\hbox{$<$}\mkern-14mu 
             \lower0.6ex\hbox{$\sim$}}}

\begin{document}

\sf 
\setlength{\textwidth}{16cm}            
\setlength{\parindent}{0mm}    

\newcommand{\aj }{\textit{The Astronomical Journal}}
\newcommand{\apj 	 }{\textit{The Astrophysical Journal}}
\newcommand{\apjl 	 }{\textit{The Astrophysical Journal, Letters}}
\newcommand{\apjlett }{\textit{The Astrophysical Journal, Letters}}
\newcommand{\apjs 	 }{\textit{The Astrophysical Journal, Supplement}}
\newcommand{\apjsupp }{\textit{The Astrophysical Journal, Supplement}}
\newcommand{\aap 	 }{\textit{Astronomy \& Astrophysics}}
\newcommand{\astap 	 }{\textit{Astronomy \& Astrophysics}}
\newcommand{\araa 	 }{\textit{Annual Review of Astronomy \& Astrophysics}}
\newcommand{\aapr 	 }{\textit{Astronomy \& Astrophysics Reviews}}
\newcommand{\aaps 	 }{\textit{Astronomy \& Astrophysics, Supplement}}
\newcommand{\mnras 	 }{\textit{Monthly Notices of the Royal Astronomical Society}}
\newcommand{\pasp 	 }{\textit{Publications of the Astronomical Society of the Pacific}}
\newcommand{\apss 	 }{\textit{Astrophysics \& Space Science}}
\newcommand{\pasj 	 }{\textit{Publications of the Astronomical Society of Japan}}
\newcommand{\nat 	 }{\textit{Nature}}
\newcommand{\sci 	 }{\textit{Science}}
\newcommand{\skytel  }{\textit{Sky \& Telescope}}
\newcommand{\memras  }{\textit{Memoirs of the Royal Astronomical Society}}
\newcommand{\qjras 	 }{\textit{Quarterly Journal of the Royal Astronomical Society}}
\newcommand{\jrasc 	 }{\textit{Journal of the Royal Astronomical Society of Canada}}
\newcommand{\jgr	 }{\textit{J. Geophys. Res.}}
\newcommand{\baas 	 }{\textit{Bulletin of the American Astronomical Society}}
\newcommand{\dps 	 }{\textit{AAS/Division for Planetary Sciences Meeting Abstracts}}
\newcommand{\iauc 	 }{\textit{IAU Circulars}}
\newcommand{\iaucirc }{\textit{IAU Circulars}}
\newcommand{\icarus	 }{\textit{Icarus}}
\newcommand{\aplett  }{\textit{Astrophysics Letters}}
\newcommand{\apspr 	 }{\textit{Astrophysics Space Physics Research}}
\newcommand{\bain 	 }{\textit{Bulletin of the Astronomical Institutes of the Netherlands}}
\newcommand{\fcp 	 }{\textit{Fundamental Cosmic Physics}}
\newcommand{\aipconf }{\textit{American Institute of Physics Conference Proceedings}}
\newcommand{\aspconf }{\textit{Astronomical Society of the Pacific Conference Series}}
\newcommand{\asslconf}{\textit{Astrophysics \& Space Science Library Conference Series}}
\newcommand{\procspie}{\textit{Proceedings of the International Society for Optical Engineering}}
\newcommand{\maps 	 }{\textit{Meteoritics \& Planetary Science}}
\newcommand{\planss  }{\textit{Planetary Space Science}}
\newcommand{\ssr 	 }{\textit{Space Science Reviews}}
\newcommand{\ao 	 }{\textit{Applied Optics}}
\newcommand{\applopt }{\textit{Applied Optics}}
\newcommand{\solphys }{\textit{Solar Physics}}
\newcommand{\pra 	 }{\textit{Physical Review A: Atomic, Molecular, and Optical Physics}}
\newcommand{\prb 	 }{\textit{Physical Review B: Condensed Matter and Materials Physics}}
\newcommand{\prc 	 }{\textit{Physical Review C: Nuclear Physics}}
\newcommand{\prd 	 }{\textit{Physical Review D: Particles, Fields, Gravitation and Cosmology}}
\newcommand{\pre 	 }{\textit{Physical Review E: Statistical, Nonlinear, and Soft Matter Physics}}
\newcommand{\prl 	 }{\textit{Physical Review Letters}}
\newcommand{\rmp 	 }{\textit{Reviews of Modern Physics}} 			
\newcommand{\grl 	 }{\textit{Geophysical Research Letters}}

\textbf{
\Large
\color{darkblue}
A ring system detected around the Centaur (10199) Chariklo \\
\normalsize
\\ Journal-ref: 	Braga-Ribas et al., Nature, Volume 508, Issue 7494, pp. 72-75 (2014)
\centerline{DOI: 10.1038/nature13155}\\
}
\vspace{0mm}

\noindent
\textbf{
F. Braga-Ribas$^{\ref{ON}}$, 
B. Sicardy$^{\ref{lesia}}$, 
J. L. Ortiz$^{\ref{iaa}}$,
C. Snodgrass$^{\ref{maxp}}$, 
F. Roques$^{\ref{lesia}}$, 
R. Vieira-Martins$^{\ref{ON},\ref{OV},\ref{imcce}}$,
J. I. B. Camargo$^{\ref{ON}}$,
M. Assafin$^{\ref{OV}}$,
R. Duffard$^{\ref{iaa}}$,
E. Jehin$^{\ref{liege}}$,
J. Pollock$^{\ref{PAD}}$,
R. Leiva$^{\ref{IApuc}}$,
M. Emilio$^{\ref{uepg}}$,
D. I. Machado$^{\ref{Foz}, \ref{Unioeste}}$, 
C. Colazo$^{\ref{MEPC}, \ref{ObsUNC}}$, 
E. Lellouch$^{\ref{lesia}}$, 
J. Skottfelt$^{\ref{NBI}, \ref{CSPF}}$, 
M. Gillon$^{\ref{liege}}$,
N. Ligier$^{\ref{lesia}}$, 
L. Maquet$^{\ref{lesia}}$, 
G. Benedetti-Rossi$^{\ref{ON}}$,
A. Ramos Gomes Jr$^{\ref{OV}}$,
P. Kervella$^{\ref{lesia}}$, 
H. Monteiro$^{\ref{ICE-MG}}$, 
R. Sfair$^{\ref{Guara}}$, 
M. El Moutamid$^{\ref{lesia},\ref{imcce}}$, 
G. Tancredi$^{\ref{Molinos},\ref{fciencias}}$,
J. Spagnotto$^{\ref{StaRosa}}$, 
A. Maury$^{\ref{SanPedro}}$,
N. Morales$^{\ref{iaa}}$,
R. Gil-Hutton$^{\ref{casleo}}$, 
S. Roland$^{\ref{Molinos}}$, 
A. Ceretta$^{\ref{fciencias},\ref{IPA}}$, 
S.-h. Gu$^{\ref{Yunnan}, \ref{ChinaAS}}$, 
X.-b. Wang$^{\ref{Yunnan}, \ref{ChinaAS}}$, 
K. Harps\o e$^{\ref{NBI}, \ref{CSPF}}$, 
M. Rabus$^{\ref{IApuc},\ref{MaxHeid} }$, 
J. Manfroid$^{\ref{liege}}$, 
C. Opitom$^{\ref{liege}}$,
L. Vanzi$^{\ref{pucChile}}$,
L. Mehret$^{\ref{uepg}}$, 
L. Lorenzini$^{\ref{Foz}}$, 
E. M. Schneiter$^{\ref{ObsUNC}, \ref{CONICET}, \ref{IATE}, \ref{UNC}}$, 
R. Melia$^{\ref{ObsUNC}}$, 
J. Lecacheux$^{\ref{lesia}}$,
F. Colas$^{\ref{imcce}}$, 
F. Vachier$^{\ref{imcce}}$,
T. Widemann$^{\ref{lesia}}$,
L. Almenares$^{\ref{Molinos},\ref{fciencias}}$, 
R. G. Sandness$^{\ref{SanPedro}}$, 
F. Char$^{\ref{UnAst}}$, 
V. Perez$^{\ref{Molinos},\ref{fciencias}}$, 
P. Lemos$^{\ref{fciencias}}$, 
N. Martinez$^{\ref{Molinos},\ref{fciencias}}$, 
U. G. J\o rgensen$^{\ref{NBI}, \ref{CSPF}}$, 
M. Dominik$^{\ref{UStAndrews}, \dag}$, 
F. Roig$^{\ref{ON}}$,
D. E. Reichart$^{\ref{UNCA}}$, 
A. P. LaCluyze$^{\ref{UNCA}}$, 
J. B. Haislip$^{\ref{UNCA}}$, 
K. M. Ivarsen$^{\ref{UNCA}}$, 
J. P. Moore$^{\ref{UNCA}}$, 
N. R. Frank$^{\ref{UNCA}}$, 
D. G. Lambas$^{\ref{ObsUNC},\ref{IATE}}$. 
}

\textit{
\vspace{-7mm}
\begin{footnotesize}
\begin{enumerate}
\setlength{\parskip}{-1mm}               
\item \label{ON} Observat\'orio Nacional/MCTI, Rua General Jos\'e Cristino 77, CEP 20921-400 Rio de Janeiro,  RJ, Brazil. 
\item \label{lesia}
LESIA, Observatoire de Paris, CNRS UMR 8109, Univ. Pierre et Marie Curie, Univ. Paris-Diderot, 5 place Jules Janssen, F-92195 MEUDON Cedex, France. 
\item \label{iaa} Instituto de Astrof\'{\i}sica de Andaluc\'{\i}a, CSIC , Apt. 3004, 18080 Granada, Spain. 
\item \label{maxp} Max Planck Institute for Solar System Research, Justus-von-Liebig-Weg 3, 37077 G\"ottingen, Germany. 
\item \label{OV} Observat\'orio do Valongo/UFRJ, Ladeira Pedro Antonio 43, CEP 20.080-090 Rio de Janeiro, RJ, Brazil. 
\item \label{imcce} Observatoire de Paris, IMCCE, UPMC, CNRS, 77 Av. Denfert-Rochereau, 75014 Paris, France. 
\item \label{liege} Institut d'Astrophysique de l'Universit\'e de Li\`ege, All\'ee du 6 Ao\^ut 17, B-4000 Li\`ege, Belgium. 
\item \label{PAD} Physics and Astronomy Department, Appalachian State Univ., Boone, NC 28608, USA.
\item \label{IApuc}  Instituto de Astrofísica, Facultad de Física, Pontificia Universidad Católica de Chile, Av. Vicuña Mackenna 4860, 
Santiago, Chile.
\item \label{uepg} Universidade Estadual de Ponta Grossa, O.A. - DEGEO, Av. Carlos Cavalcanti 4748, Ponta Grossa 84030-900, PR, Brazil. 
\item \label{Foz} Polo Astron\^omico Casimiro Montenegro Filho / FPTI-BR, Av. Tancredo Neves, 6731, CEP 85867-900, Foz do Igua\c cu, PR, Brazil.
\item \label{Unioeste} Universidade Estadual do Oeste do Paran\'a (Unioeste), Av. Tarqu\'inio Joslin dos Santos, 1300, CEP 85870-650, Foz do Igua\c cu, PR, Brazil.
\item \label{MEPC} Ministerio de Educaci\'on de la Provincia de C\'ordoba, C\'ordoba, Argentina.
\item \label{ObsUNC} Observatorio Astron\'omico, Universidad Nacional de C\'ordoba, C\'ordoba, Argentina. 
\item \label{NBI} Niels Bohr Institute, University of Copenhagen, Juliane Maries vej 30, 2100 Copenhagen, Denmark.
\item \label{CSPF} Centre for Star and Planet Formation, Geological Museum, \O ster Voldgade 5, 1350 Copenhagen, Denmark.
\item \label{ICE-MG}  Instituto de Física e Química, Av. BPS 1303, CEP 37500-903, Itajub\'a, MG, Brazil.
\item \label{Guara} UNESP - Univ Estadual Paulista, Av Ariberto Pereira da Cunha, 333, CEP 12516-410 Guaratinguet\'a, SP, Brazil. 
\item \label{Molinos} Observatorio Astronomico Los Molinos, DICYT, MEC, Montevideo, Uruguay.
\item \label{fciencias} Dpto. Astronomia, Facultad Ciencias, Uruguay.
\item \label{StaRosa} Observatorio El Catalejo, Santa Rosa, La Pampa, Argentina.
\item \label{SanPedro} San Pedro de Atacama Celestial Explorations, Casilla 21, San Pedro de Atacama, Chile.
\item \label{casleo} Complejo Astron\'omico El Leoncito (CASLEO) and San Juan National University, Av. Espa\~na 1512 sur, J5402DSP, San Juan, Argentina. 
\item \label{IPA} Observatorio del IPA, Consejo de Formaci\'on en Educaci\'on, Uruguay.
\item \label{Yunnan} Yunnan Observatories, Chinese Academy of Sciences, Kunming 650011, China.
\item \label{ChinaAS} Key Laboratory for the Structure and Evolution of Celestial Objects, Chinese Academy of Sciences, Kunming 650011, China.
\item \label{MaxHeid} Max Planck Institute for Astronomy, K\"onigstuhl 17, 69117 – Heidelberg, Germany.
\item \label{pucChile} Department of Electrical Engineering and Center of Astro-Engineering, Pontificia Universidad Cat\'olica de Chile, Av. Vicu\~na Mackenna 4860, Santiago, Chile.
\item \label{CONICET} Consejo Nacional de Investigaciones Científicas y Técnicas (CONICET), Argentina.
\item \label{IATE} Instituto de Astronom\' ia Te\'orica y Experimental IATE–CONICET, C\'ordoba, Argentina.
\item \label{UNC} Facultad de Ci\^ encias Exactas, F\'isicas y Naturales, Universidad Nacional de C\'ordoba (UNC), C\'ordoba, Argentina.
\item \label{UnAst} Unidad de Astronomía, Facultad de Ciencias Básicas, Universidad de Antofagasta, Avenida Angamos 601, Antofagasta, Chile.
\item \label{UStAndrews} SUPA, University of St Andrews, School of Physics \& Astronomy,
North Haugh, St Andrews, KY16 9SS, United Kingdom. $^\dag$Royal Society University Research Fellow. 
\item \label{UNCA} Department of Physics and Astronomy, University of North Carolina - Chapel Hill, North Carolina.
\end{enumerate}
\end{footnotesize}
}
  
\setlength{\topmargin}{-2.cm}            
\setlength{\textwidth}{16cm}            	
\setlength{\textheight}{24cm}		
\setlength{\parindent}{4mm}

\newpage


\textbf{
Until now, rings have been detected in the Solar System exclusively around the four giant planets\cit{Tiscareno13}.
%
Here we report the discovery of the first minor-body ring system around the Centaur object (10199) Chariklo,
a body with equivalent radius\cit{Fornasier13} 124$\pm$9~km. 
%
A multi-chord stellar occultation 
revealed the presence of two dense rings around Chariklo, 
with widths of about 7~km and 3~km, optical depths 0.4 and 0.06, 
and orbital radii 391 and 405~km, respectively.
%
The present orientation of the ring is consistent with an edge-on geometry in 2008, 
thus providing a simple explanation 
for the dimming of Chariklo's system between 1997 and 2008\cit{Belskaya10}, 
and for the gradual disappearance of ice and other absorption features in its
spectrum over the same period\cit{Guilbert09,Guilbert11}. 
This implies that the rings are partially composed of water ice. 
%
These rings may be the remnants of a debris disk, which were possibly confined by embedded kilometre-sized satellites. 
%
%
%
}



Chariklo is the largest known Centaur object
orbiting in a region between Saturn and Uranus 
with orbital eccentricity 0.171 and semi-major axis 15.8 astronomical units (1AU is the Earth-Sun distance).  
It may be a former transneptunian object that has been recently (less than 10 Myr)
scattered by gravitational perturbations from Uranus\cit{Horner04}. 
No clear detection of Chariklo’s rotation has been made so far.
Its surface is very dark with a geometric albedo\cit{Fornasier13} 0.035$\pm$0.011, and
it is subject to long-term spectral and photometric variabilities\cit{Belskaya10,Guilbert09,Guilbert11}, 
although no cometary activity has ever been reported.

An occultation of an R=12.4 magnitude star by Chariklo was predicted\cit{Camargo13} 
to cross South America on June 3, 2013; see Extended Data Figures~\ref{fig_map}~and~\ref{StarSpec}. 
We obtained data from sites distributed in Brazil, Argentina, Uruguay and Chile (Extended Data Table~\ref{tab_obs}).  
While the occultation by Chariklo itself was recorded at three sites in Chile, seven sites detected a total of thirteen
rapid stellar flux interruptions (secondary events), two of them being resolved into two sub-events 
by the Danish 1.54m telescope at the European Southern Observatory (ESO) at La Silla (Figure~\ref{fig_DanishCurve}).

\begin{figure}[!h]
\centerline{\includegraphics[totalheight=12cm,trim=0 30 0 50,angle=0]{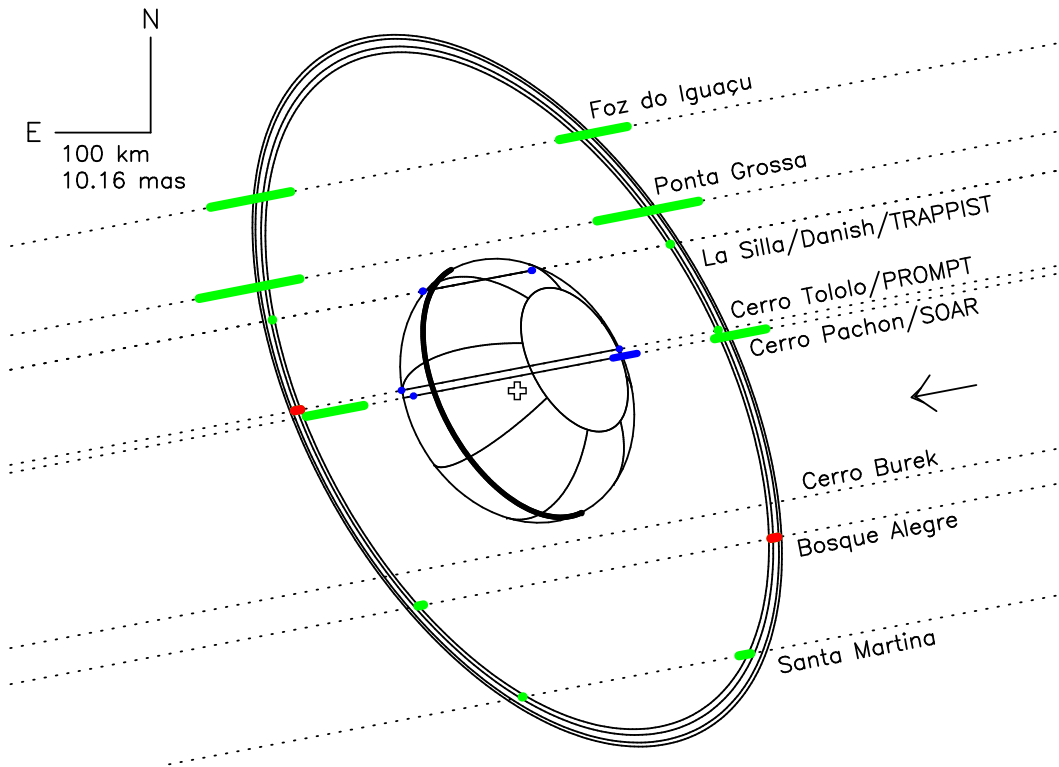}}
\caption{ 
\sf \textbf{Figure~\ref{fig_rings} $|$ Chariklo ring system.}
The dotted lines are the trajectories of the star relative to Chariklo
in the plane of the sky, as observed from eight sites (see the SI for details),
the arrow indicating the direction of motion.
The green segments represent the locations of ring C1R observed at each station 
(at 1$\sigma$-level uncertainty).
For clarity, we have not plotted the detections made at the TRAPPIST and 0.275~m 
telescopes (at La Silla and Bosque Alegre, respectively) as they have larger error bars
than their local counterparts, and would supersede the corresponding green segments.
Two ring events occurred during camera readout times (red segments), 
at Bosque Alegre and Cerro Tololo, and also provide constraints on the ring orbit. 
The ring events are only marginally detected at Cerro Burek, but the signal-to-noise ratio 
is not sufficient to bring further constraints on the ring orbit and equivalent width.
An elliptical fit to the green and red segments (excluding the SOAR events at Cerro Pach\'on due
to timing problems, see the SI) 
provides the centre of the rings (see open cross), as well as their sizes, opening angle and orientation, 
see Table~\ref{tab_ring_prop}.
Chariklo's limb has been fitted to the two chords extremities (blue segments) obtained at La Silla and Cerro Tololo, 
assuming that the centres of Chariklo and the rings coincide as well as their position angles. 
This assumption is expected if Chariklo is a spheroid, with a circular ring orbiting in the equatorial plane, 
see text and the SI. }
\label{fig_rings}
\end{figure}

Displayed in the Extended Data (ED) and analysed in the Supplementary Information (SI), all those secondary events (see ED Tables~\ref{tab_ring_timings}~and~\ref{tab_ring_danish}) can be readily explained 
by the presence of two narrow and azimuthally homogeneous rings (Figure~\ref{fig_rings})
whose widths and optical depths are given in Table~\ref{tab_ring_prop}.
Even if the events were generally not resolved (see ED Fig.~\ref{fig_all_fits}), their depths provide a measure of the integrated light loss of the events,
which in turn depends on the local geometry of the occultation  in the plane of the sky. 
The fact  that all the events are consistent with an azimuthally homogeneous ring system makes  other interpretations, such as an ensemble of cometary jets, very unlikely.

Other evidence supports our interpretation of a flat circular ring system around Chariklo. 
The ellipse fitted to the secondary events provides two possible ring pole positions (ED Table~\ref{tab_ring_geo}).
Our preferred solution is the one where the rings had 
an opening angle of  60\dg\ in 1996--1997, and vanished out of view as they were observed edge-on in 2008, due to the orbital motion of Chariklo relative to the Earth.
It provides a simple explanation for the gradual dimming of Chariklo's system, by a factor of 1.75, during that period\cit{Belskaya10}.
Further evidence is that the 2~$\mu$m water ice band and the spectral slope below 0.55~$\mu$m,
gradually disappeared\cit{Guilbert09,Guilbert11} between 1997 and 2008, implying that water ice is present in the rings. 
Observations made in 2013 show that the system brightened by a factor of about 1.5 since 2008, and that the water ice band is detectable again,  supporting our interpretation (R.D. \etal, in preparation).

\begin{figure}[!h]
\centerline{\includegraphics[totalheight=12cm,trim=0 30 0 50,angle=0]{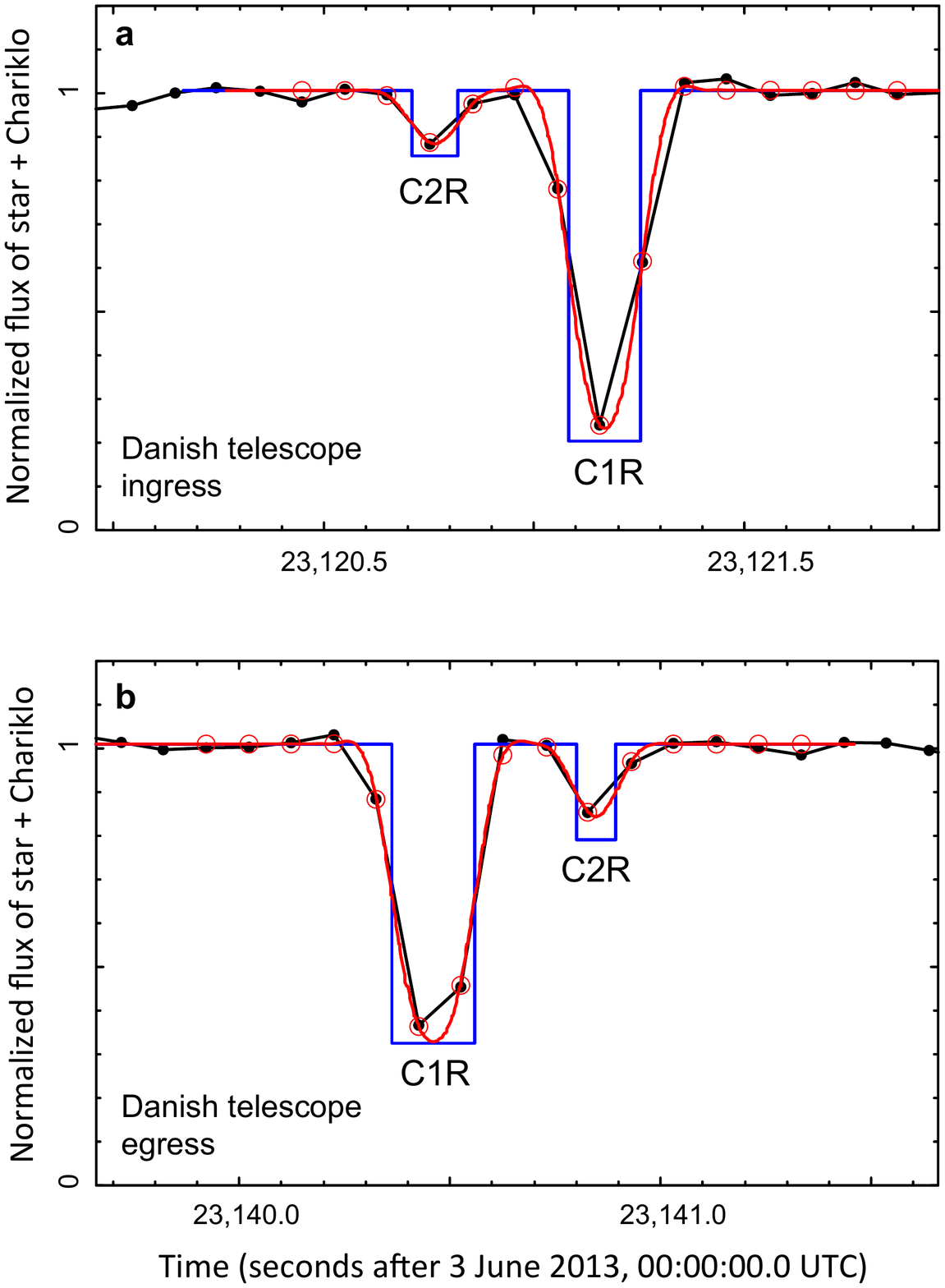}}
\caption{ 
\sf \textbf{Figure~\ref{fig_fit_danish} $|$ Fits to the Danish ring events.}
The red curves are synthetic occultation profiles produced by semi-transparent bands 
with square-well profiles (the blue lines), after convolution by 
Fresnel diffraction, observed bandwidth, 
the stellar radius projected at Chariklo, and 
the finite integration time.
The open red circles are the values of the model for the times corresponding to the observed data points
at ingress and egress (panels \textbf{a} and \textbf{b}, respectively)
The $\chi^2$ values per degree of freedom of the fits to the four ring events  
vary from 0.4 to 1.2, see the Extended Data Table~\ref{tab_ring_timings}. 
This indicates satisfactory fits, and shows that the events are compatible with sharp-edged rings.
The resulting widths and optical depths of rings C1R and C2R are listed in Table~\ref{tab_ring_prop},
after the appropriate projections into the plane of the rings have been performed. 
Examination of ED Table~\ref{tab_ring_danish} shows that the 
widths and optical depths of ring C1R at the Danish 1.54~m telescope 
differ moderately but significantly between ingress and egress. 
The equivalent depth of ring C1R differs by 21\% between ingress and egress. 
Similar variations are observed in Uranus' narrow rings, and might be associated with normal 
mode oscillations that azimuthally modulates the width and optical depth of the rings\cit{French91}. Differences between C2R ingress/egress are marginally significant.
}%
\label{fig_fit_danish}
\end{figure}

Due to its higher acquisition rate (10Hz), 
the Lucky Imager camera\cit{Harpsoe12,Skottfelt13} of the Danish 1.54~m telescope
actually resolved the secondary events into two rings,
denoted  ``2013C1R" and ``2013C2R" (``C1R'' and ``C2R'' for short) in Fig.~\ref{fig_fit_danish}.
Hereby, the term ``ingress" (resp. ``egress") refers to the first (resp. second) of a pair of ring events at a given site.
All the Danish events are satisfactorily fitted by sharp-edged ring models
whose radial widths ($W$), in the ring plane, and normal optical depths 
($\tau_{\sf _{N}}$), are listed in Table~\ref{tab_ring_prop}.
We also provide the equivalent depths ($E_\tau= W \tau_{_{N}}$),
which can be related to the amount of material contained in the ring\cit{French91}.
The ring C2R is about 40\% narrower than C1R, and contains about 12 times less material.
Note that no material is detected in the
gap between C1R and C2R, up to a limit of 0.004 in normal optical depth and 0.05~km 
in equivalent depth (ED Table~\ref{tab_ring_geo}). 

By analogy with Saturn's A ring\cit{Colwell09} or the dense Uranus' rings\cit{Esposito91}, 
we estimate that the surface density of C1R lies in the range 30~--~100~g~cm$^{-2}$ (see SI). 
Then, the mass of C1R is equivalent to an icy body of radius of roughly one kilometre, 
while the ring C2R would correspond to a body of half that size.
If the photometric variability of Chariklo's system\cit{Belskaya10}  
between 1997 and 2008 
is entirely due to the ring changing geometry,
we estimate the ring reflectivity $I/F~\sim~0.09~\pm~0.04$ 
($I$ is the intensity of light reflected by the surface, $\pi F$ is the incident solar flux density).
Thus, Chariklo's ring particles would be significantly 
brighter than those of Uranus' rings\cit{Karkoschka01} ($I/F~\sim~0.05$),
but significantly darker than those of Saturn's A ring\cit{Hedman13} ($I/F~\sim~0.3$).
Note that, if part of the photometric variability is caused by Chariklo itself, 
then the ring 
material would be darker than estimated above; see SI.

Constraints on Chariklo's limb shape are based on only two occultation chords (see SI and ED Table \ref{tab_Chariklo_timings}).
Our simplest model describes an oblate Chariklo surrounded by a circular equatorial ring system (see ED Table~\ref{tableChar}). 
%
%
The fitted limb (Fig. \ref{fig_rings}) has an equivalent radius of 127~km 
(the radius of an equivalent spherical body), 
consistent with the value derived from thermal data\cit{Fornasier13}, 124~$\pm$~9~km, thus supporting our model.

From the Roche critical density limit\cit{Tiscareno13b}, we estimate in the SI that typical densities for Chariklo and ring particles, are consistent to explain the present rings location.
Moreover, an unperturbed ring of some kilometres in width
and thickness $h$ should spread, due to inter-particle collisions\cit{Goldreich79}, 
in $10^{4}/(h/\sf{metre})^{2} \sim$ a few thousand years, assuming 
$h$ of a few metres (by analogy with Saturn's rings).
Furthermore, Poynting-Robertson drag\cit{Goldreich79} should spread sub-cm particles
in a few million years at most (see SI).
Thus, the rings are either very young or actively confined. A confinement mechanism, may be provided by kilometre-sized ``shepherd satellites" that would have a mass comparable to that of the rings, see the SI.

We do not know if rings around minor bodies stem from a generic, yet unknown process, or are exceptional systems.
We note that many stellar occultations by main belt asteroids and almost ten transneptunian 
events\cit{Elliot10,Sicardy11,Ortiz12,BragaRibas13}
did not reveal rings so far (nor did direct images).
Stellar occultations and appulses involving (2060) Chiron (another Centaur object similar in size to Chariklo) 
in 1993 and 1994, revealed a narrow jet-like feature and diffuse material around that object\cit{Bus96,Elliot95}.
This was interpreted as material ejected from the surface, supported by the fact that
Chiron is known to be an active comet-like object.
It is unclear if the detection of material around both objects is a mere
coincidence, or if they share a common physical process (noting that no cometary activity has been detected around Chariklo).

About 5\% of the Centaur and transneptunian population\cit{Brunini14} are known to have satellites.
While the large satellites are thought to result from three-body captures, 
their small counterparts are more likely to form from impacts\cit{Noll08}, or rotational disruptions\cit{Ortiz12b}, possibly re-accreted from a remaining disk after that.
To date, no observations have shown satellites around Chariklo
(the rings span at most 0.04~arcsec around the primary, 
making direct detections of associated small satellites a challenge).
Several scenarios for Chariklo's rings origin can be proposed,
all relying on a disk of debris where the largest elements acted as shepherds 
for the smaller material: 
(1) an impactor that excavated icy material from Chariklo's outer layers; or destroyed a pre-existing satellite; or was itself disrupted during the impact; 
(2) a disk of debris formed from a rotational disruption of the main body or fed by a cometary-like activity;
(3) two pre-existing satellites that collided through a mechanism yet to be explained 
or
(4) a retrograde satellite that migrated inward and eventually got disrupted by tidal forces.

Note that the mass and angular momentum of the rings and their hypothetical shepherds 
are very small (by a factor of less than $10^{-5}$) compared to that of Chariklo.
The typical escape velocity at the surface of Chariklo is $\sim$~0.1~km~s$^{-1}$.
Thus, if an impact from an outsider generated the rings, it must involve low incoming velocities.
While the impact velocities in the Main Belt of asteroids are of the order of 5~km~s$^{-1}$, 
they are $\sim$~1~km~s$^{-1}$ in the outer Solar System, and even lower before the Kuiper Belt was dynamically excited\cit{Cuk13},
possibly explaining why no rings have been found yet around main belt asteroids.
Finally, Chariklo's orbit is perturbed by Uranus, which transferred  the Centaur 
from the transneptunian region less than $\sim$~10~Myr ago\cit{Horner04}.
As estimated in the SI, a very close encounter at about five Uranus radii is actually
necessary to disrupt the ring system. 
Such an event has a small probability of occurrence\cit{Nogueira11}, which supports
the possibility that the rings formed in the transneptunian region and
survived the transfer episode.


\setlength{\parskip}{0mm}
\setlength{\parindent}{-6mm}

\setlength{\parskip}{4mm}
\setlength{\parindent}{0mm}
 
{\bf Supplementary Information} is linked to the online version of the paper at \\ \url{www.nature.com/nature}.


\noindent
\textbf{Acknowledgements.}
We thank  S. Fornasier, I. Belskaya, B. Carry, E. Nogueira, P. Michel and A. Morbidelli
for the discussions that helped to interpret our results. The authors thank Joseph A. Burns and the anonymous referee for their comments which helped to improve the paper.
F.B.R. acknowledges LIneA (Laborat\'orio Interinstitucional de e-Astronomia) for hosting the campaign web-page and the support (grant 150541/2013-9) of CNPq, Brazil.
Operation of the Danish 1.54m telescope is financed by a grant to U.G.J. by 
the Danish Natural Science Research Council (FNU) and by the Centre for Star and Planet Formation (StarPlan).
F.C., J.L., L.M., F.R., B.S., F.V. and T.W. acknowledge support from the French grant `Beyond Neptune II'.  
J.L.O., R.D. and N.M. acknowledge funding from Spanish AYA grants and FEDER funds. 
TRAPPIST is a project funded by the Belgian Fund for Scientific Research (FRS-FNRS)
with the participation of the Swiss National Science Fundation (SNF). E. J. and M. G. are FNRS Research Associates, J. M. is Research Director FNRS. C. O. thanks the Belgian FNRS for funding her PhD thesis.
J.I.B.C., M.A. and R.V.M. acknowledges CNPq grants 302657/2010-0, 482080/2009-4 478318/2007-3 and 304124/2007-9, respectively.
The Universidad Cat\'olica Observatory (UCO) Santa Martina is operated by the Pontif\'icia Universidad Cat\'olica de Chile (PUC).
C.S. received funding from the European Union Seventh Framework Programme (FP7/2007-2013) 
under grant agreement no. 268421. 
S.h.G. and X.b.W. thank the financial support from National Natural Science Foundation of China through grants 10873031 and 11073051.
M.R. acknowledges support from FONDECYT postdoctoral fellowship 3120097.
Partially based on observations made at the Pico dos Dias Observatory from the National Laboratory of Astrophysics (OPD/LNA) – Brazil.
Partially based on observations obtained at the Southern Astrophysical Research (SOAR) telescope, 
which is a joint project of the Minist\'{e}rio da Ci\^{e}ncia, Tecnologia, e Inova\c{c}\~{a}o (MCTI) 
da Rep\'{u}blica Federativa do Brasil, the U.S. National Optical Astronomy Observatory (NOAO), 
the University of North Carolina at Chapel Hill (UNC), and Michigan State University (MSU).
This publication makes use of data products from the Two Micron All Sky Survey, 
a joint project of the University of Massachusetts and the Infrared Processing and Analysis Center/California Institute of Technology, 
funded by the National Aeronautics and Space Administration (NASA) 
and the National Science Foundation. It also makes use of data products from the 
Wide-field Infrared Survey Explorer, which is a joint project of the 
University of California, Los Angeles, and the Jet Propulsion Laboratory/California Institute of Technology, funded by NASA.
UNC-CH gratefully acknowledges NSF Awards 0959447, 1009052, and 1211782 for support of Skynet/PROMPT.
L.V. and R.L. acknowledge support by CONICYT through the project Anillo ACT-86.
A.M., acknowledge the use of Caisey Harlingten's 20 inch Planewave telescope, 
which is part of the Searchlight Observatory Network.


\noindent
\textbf{Author contributions.}
F.B.R. planned the observation campaign, centralised the occultation predictions, participated in the observations, analysed data and results, ran the diffraction and limb-fitting codes and wrote the paper.
B.S. helped to plan the campaign, analysed data and results, wrote and ran the diffraction and limb-fitting codes, and wrote the paper.
J.L.O. helped to plan the campaign, analysed data for the prediction, obtained and analysed data and results.
C.S. coordinated and analysed the data from Danish 1.54~m telescope.
F.R. analysed data, wrote and ran the diffraction and limb-fitting codes.
R.V.M. coordinated the predictions, analysed data.
J.I.B.C., F.B.R., R.V.M. and M.A. discovered the star candidate and analysed data for the prediction.
R.D. coordinated the SOAR and Bosque Alegre observations and helped to analyse the results.
E.J., J.P, R.L., M.E., D.I.M., C.C., J.S., obtained and analysed the positive occultation detections from TRAPPIST, PROMPT, UCO Sta. Martina, UEPG, Foz do Igua\c cu, Bosque Alegre and Danish telescopes, respectively.
E.L. helped to analyse the results.
J.S., M.G., L.M., N.L., G.B.R., A.R.G., helped to obtain the occultation light curves and/or analysed the data.
P.K. ran stellar models to determine the apparent star diameter.
H.M. obtained the stellar spectrum.
R.S. calculated the rings stability in short term.
M.E.M. worked on the confinement mechanisms of the rings.
G.T., J.P., A.M., N.M., R.G.H., S.R., A.C., obtained the negative detections.
S.h.G., X.b.W., K.H., M.R., J.M., C.O., L.V., L.M., L.L., E.M.S., R.M.,  helped to obtain and analyse the positive occultation detections.
J.L., F.C., F.V., T.W., K.H., helped to make the predictions and analyse the data.
L.A., B.S., F.C., V.P., P.L., N.M., U.G.J., M.D., F.R., D.E.R., A.P.L., J.B.H., K.M.I., J.P.M., N.R.F., D.G., are related to the event observations.
All authors were given the opportunity to review the results and comment on the manuscript.

\noindent
\textbf{Author information.}
Correspondence and requests for materials should be addressed to Felipe Braga-Ribas (ribas@on.br).\\

\clearpage

\setlength{\topmargin}{-3.5cm}            
\setlength{\textwidth}{16cm}            	
\setlength{\textheight}{26cm}		
\setlength{\parindent}{4mm}		
\setlength{\parskip}{3mm}                    

\noindent

\begin{table}
\begin{center}
\sf
\footnotesize
\setlength{\tabcolsep}{1mm}
\renewcommand{\arraystretch}{1.5} 
\begin{tabular}{llll}
\hline
\hline
       & radial width & normal optical depth & equivalent depth\\
       &   $W$  (km) & $\tau_N$                     &  $E_\tau= W \cdot \tau_N$ (km)  \\
\hline
ring C1R, Danish ingress &  6.16 $\pm$ 0.11 & 0.449 $\pm$ 0.009 & 2.77 $\pm$ 0.04  \\
ring C1R, Danish egress  &  7.17 $\pm$ 0.14 & 0.317 $\pm$ 0.008 & 2.28 $\pm$ 0.03  \\
\hline
ring C2R, Danish ingress & 3.6 $\sf ^{+1.3}_{-2.0}$ & 0.05 $\sf ^{+0.06}_{-0.01}$ & 0.18 $\pm$ 0.03  \\
ring C2R, Danish egress  & 3.4 $\sf ^{+1.1}_{-1.4}$ & 0.07 $\sf ^{+0.05}_{-0.03}$ & 0.24 $\pm$ 0.02  \\
\hline
\hline 
Ring C1R radius (km)                                                 & & 390.6 $\pm$ 3.3 & \\ 
Ring C2R radius (km)                                                 & & 404.8 $\pm$ 3.3 & \\ 
 \multicolumn{2}{l}{Radial separation of rings C1R and C2R} & 14.2 $\pm$ 0.2 km            \\
 \multicolumn{2}{l}{Gap between rings C1R and C2R}    &  
 \multicolumn{2}{l}{8.7 $\pm$ 0.4 km ($\tau_N <$ 0.004, $E_\tau <$ 0.05~km)}       \\
Opening angle $B$  (deg)                                                     & &   $B$= +33.77 $\pm$ 0.41    &   \\
Position angle $P$   (deg)                                                     & &  
\multicolumn{2}{l}{$\bf P$= \textbf{-61.54 $\pm$ 0.14 (preferred)}  or $118.46\pm0.14$}    \\
\hline
\hline
 & \multicolumn{3}{l}{Pole position (equatorial J2000)} \\
                                                  & & \textbf{solution 1 (preferred)} & solution 2 \\
\hline
Right ascension    (deg)        & & $\bf \alpha_{p}$ \textbf{= 151.30 $\pm$ 0.49}   & $\alpha_{\sf p}$ = 26.96 $\pm$ 0.29  \\
Declination             (deg)       & & $\bf \delta_{p}$ \textbf{= 41.48 $\pm$ 0.21}      &  $\delta_{\sf p}$ =  03.44 $\pm$ 0.31   \\
\hline
\hline
\end{tabular}
\end{center}
\normalsize \sf
\caption{  
\textbf{Table \ref{tab_ring_prop} $|$ Ring physical parameters.} 
\label{tab_ring_prop}
The geocentric distance of Chariklo's system at the moment of the occultation, 
$D = 2.031 \times 10^{9}$~km, provides a scale of 9846~km per arcsecond in the plane of the sky.
All the secondary events besides those of the Danish 1.54m telescope (Fig.~\ref{fig_fit_danish}) 
are satisfactorily fitted by a model where the rings C1R and C2R have widths and optical depths
that are averages of the Danish ingress and egress values given above, i.e. 
$W_{\sf C1R}= 6.6$~km, $\tau_{\sf N,C1R}= 0.38$,
$W_{\sf C2R}= 3.4$~km and $\tau_{\sf N,C2R}= 0.06$, 
see the SI.
The ring opening angle $B$ is the absolute value of the elevation of the observer above the ring plane.
The position angle $P$  is the angle between the semi-minor axis of the ring  projected in the plane of the sky, 
counted positively from celestial north to celestial east. By convention, it refers to the projected semi-major axis 
that corresponds to superior conjunction. 
The solution that best explains the photometric and spectral variations of 
Chariklo's system\cit{Belskaya10,Guilbert09,Guilbert11}  is chosen as preferred (see text).
}%
\end{table}


\clearpage 


\begin{figure}[!h]
\centerline{\includegraphics[totalheight=12cm,trim=1.5cm 0cm 3cm 3cm, clip,angle=0]{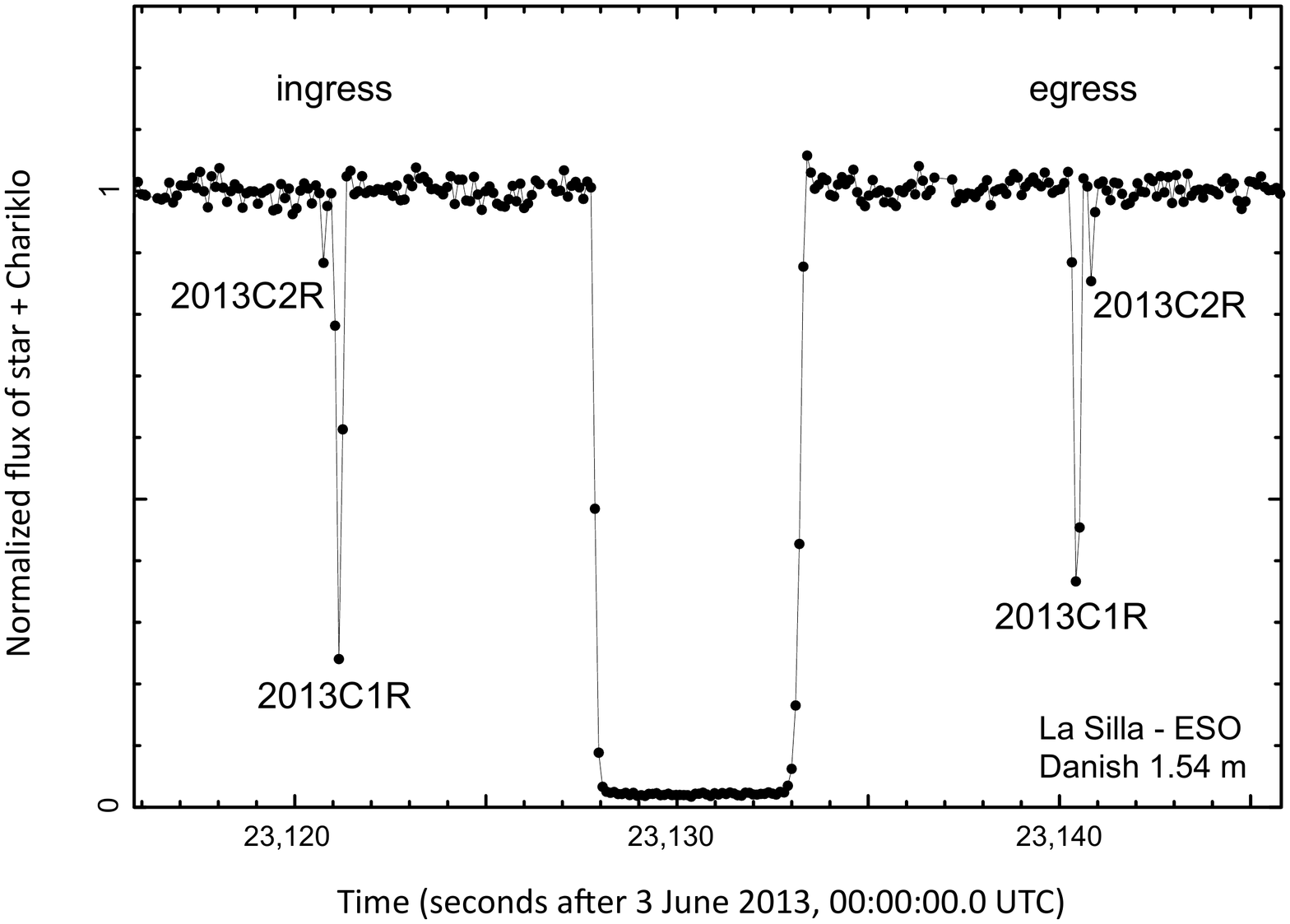}}
\caption{ 
\sf \textbf{Figure~\ref{fig_DanishCurve} $|$ Light curve of the occultation by Chariklo system.}
The data were taken with the Danish 1.54~m telescope (La Silla) on 3 June 2013, at a rate of almost 10~Hz in a long-pass filter and cut-off below 650 nm, 
limited in the red by the sensitivity of the CCD chip, see SI for more details.
Aperture photometry provided the flux from the target star and a fainter nearby reference star.  
Low-frequency sky-transparency variations were removed by dividing the target flux by 
an optimal running average of 87 data points (8.7~s) of the reference star, resulting in a final SNR of 64 per data point.
The sum of the star and Chariklo fluxes has been normalized to unity outside the occultation.
The central drop is caused by Chariklo, and two secondary events 2013C1R and 2013C2R 
are observed at ingress (before the main Chariklo occultation) and egress (after the main occultation).
A more detailed view of these ring events is shown in Fig.~\ref{fig_fit_danish}. }%
\label{fig_DanishCurve}
\end{figure}


\vspace{-6mm}

\setcounter{figure}{0}
\setcounter{table}{0}
\setcounter{equation}{0}

\fontsize{8pt}{8pt}
\setlength{\topmargin}{-2.cm}            
\setlength{\textwidth}{16cm}            	
\setlength{\textheight}{24cm}		
\setlength{\parindent}{4mm}



\begin{figure}[!h]
\centerline{\includegraphics[totalheight=7.5cm,trim=3cm 3cm 3cm 3cm, clip, angle=0]{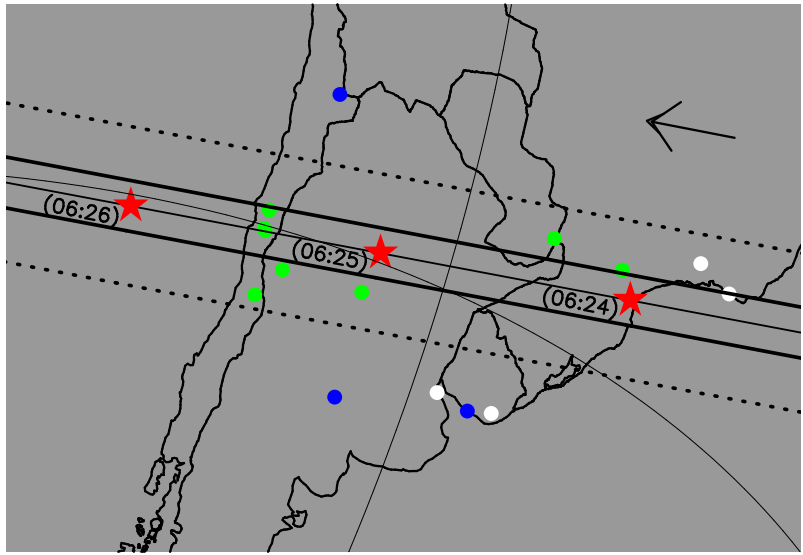}}
\caption{ 
\normalsize
\sf
\textbf{Extended Data Figure \ref{fig_map} $|$ The Chariklo 03 June 2013 occultation campaign.}
\label{fig_map} 
The continuous straight lines indicate Chariklo's shadow track on Earth, while the dotted lines correspond 
to the ring shadow, as reconstructed from our post-occultation  analysis.
The shadows move from right to left, see the arrow. 
The red stars indicate the center of Chariklos's shadow at various UT times. 
The green dots are the sites where the occultation was detected. The blue dots are the sites that had obtained data but did not detect the event, and the white ones are the sites that were clouded out (Extended Data Table~\ref{tab_obs}). 
}
\end{figure}


\begin{figure}[!h]
\centerline{\includegraphics[totalheight=7cm,trim=0 0 0 00,angle=0]{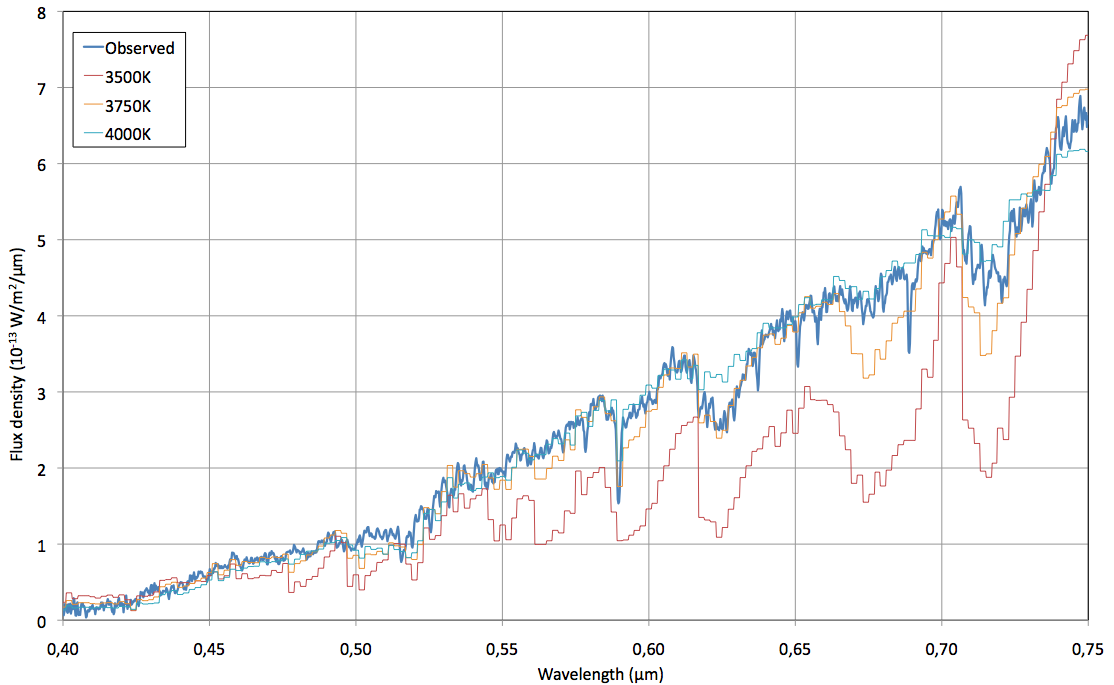}}
\caption{ 
\normalsize
\sf
\textbf{Extended Data Figure \ref{StarSpec} $|$ The occulted star spectrum and model.}
The star spectrum obtained at Pico dos Dias Observatory/Brazil with the 1.6~m telescope and a Cassegrain spectrograph. Observations were made with the spectrograph configuration, used a 600~l/mm grating, giving a resolution of 2.3~$\sf \AA$/pixel covering from 3700 to 7700~$\sf \AA$. We obtained three 300~s exposures with a 2" slit. The calibration was done with the usual procedures and one flux standard star (LTT6248).
The thick blue curve represents the observed spectrum. The red, orange and blue thin curves represent model spectra database\cit{CastelliKurucz03} used for comparison (see text).
}
\label{StarSpec}
\end{figure}


\begin{figure}[!h]
\centerline{
\includegraphics[totalheight=9.2cm,trim=11mm 1.7cm 2.9cm 4.1cm, clip]{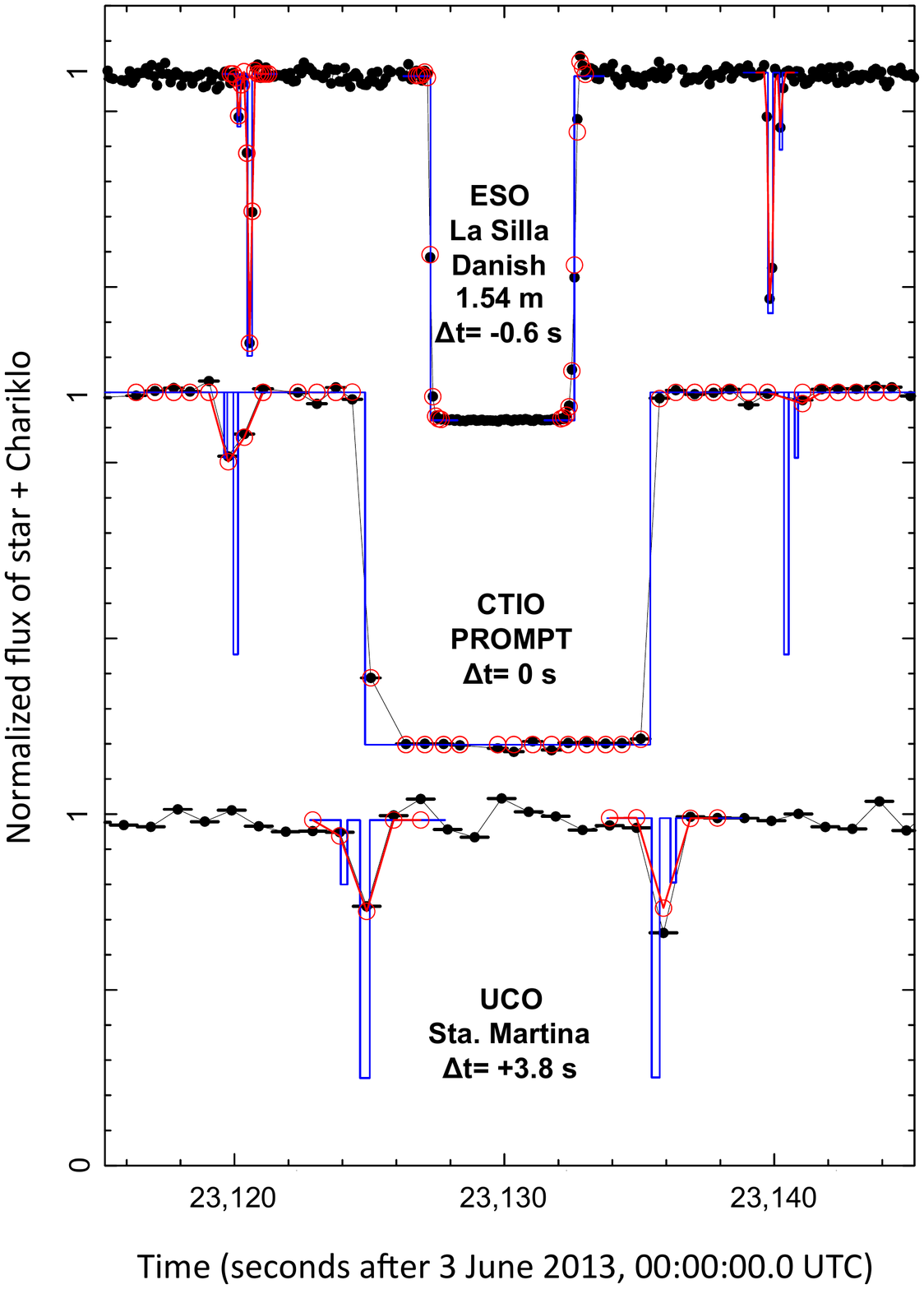} ~\ 
\includegraphics[totalheight=9.2cm,trim=11mm 1.7cm 2.9cm 4.1cm, clip]{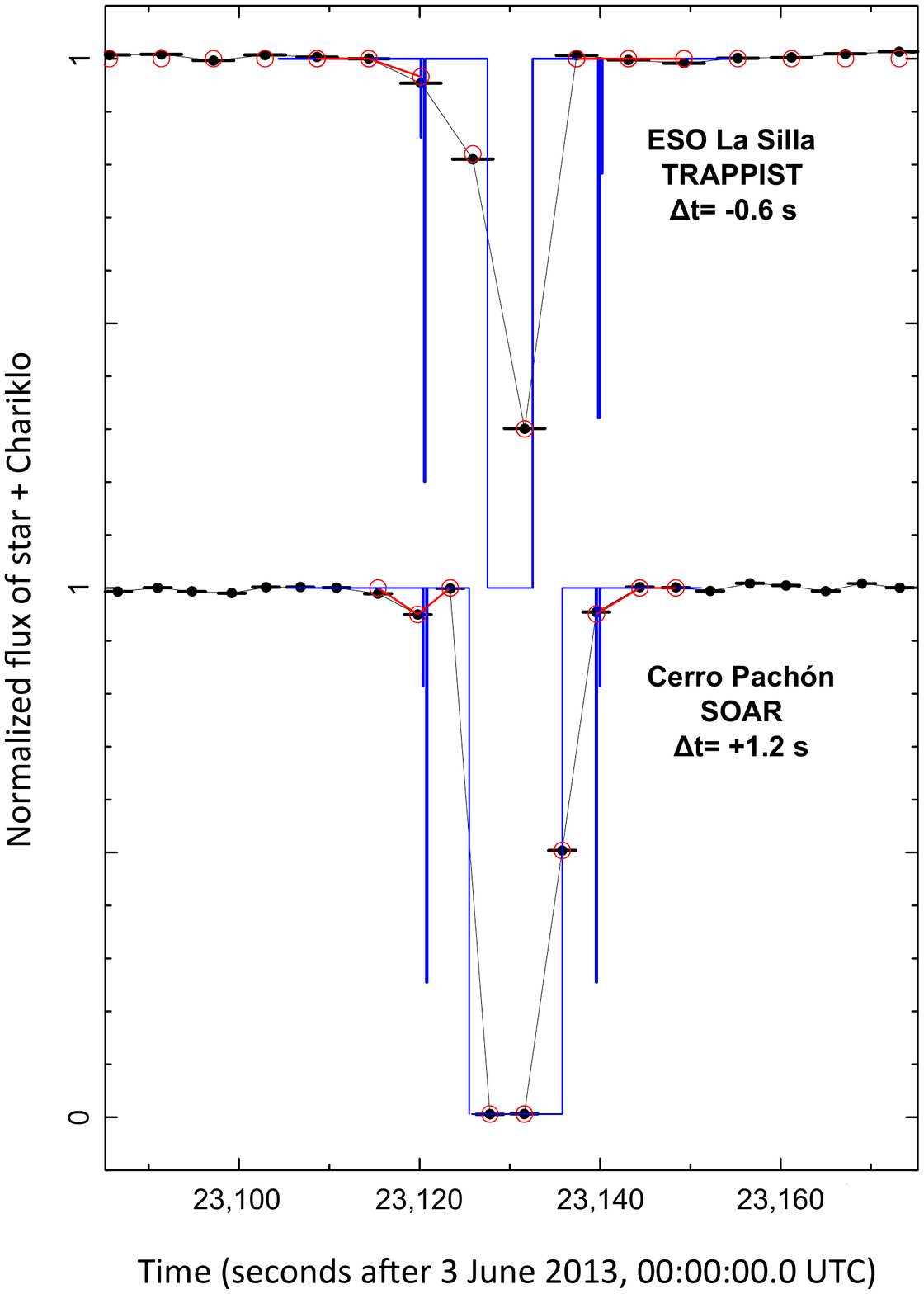}
}
\centerline{
\includegraphics[totalheight=9.2cm,trim=11mm 1.7cm 2.9cm 4.1cm, clip]{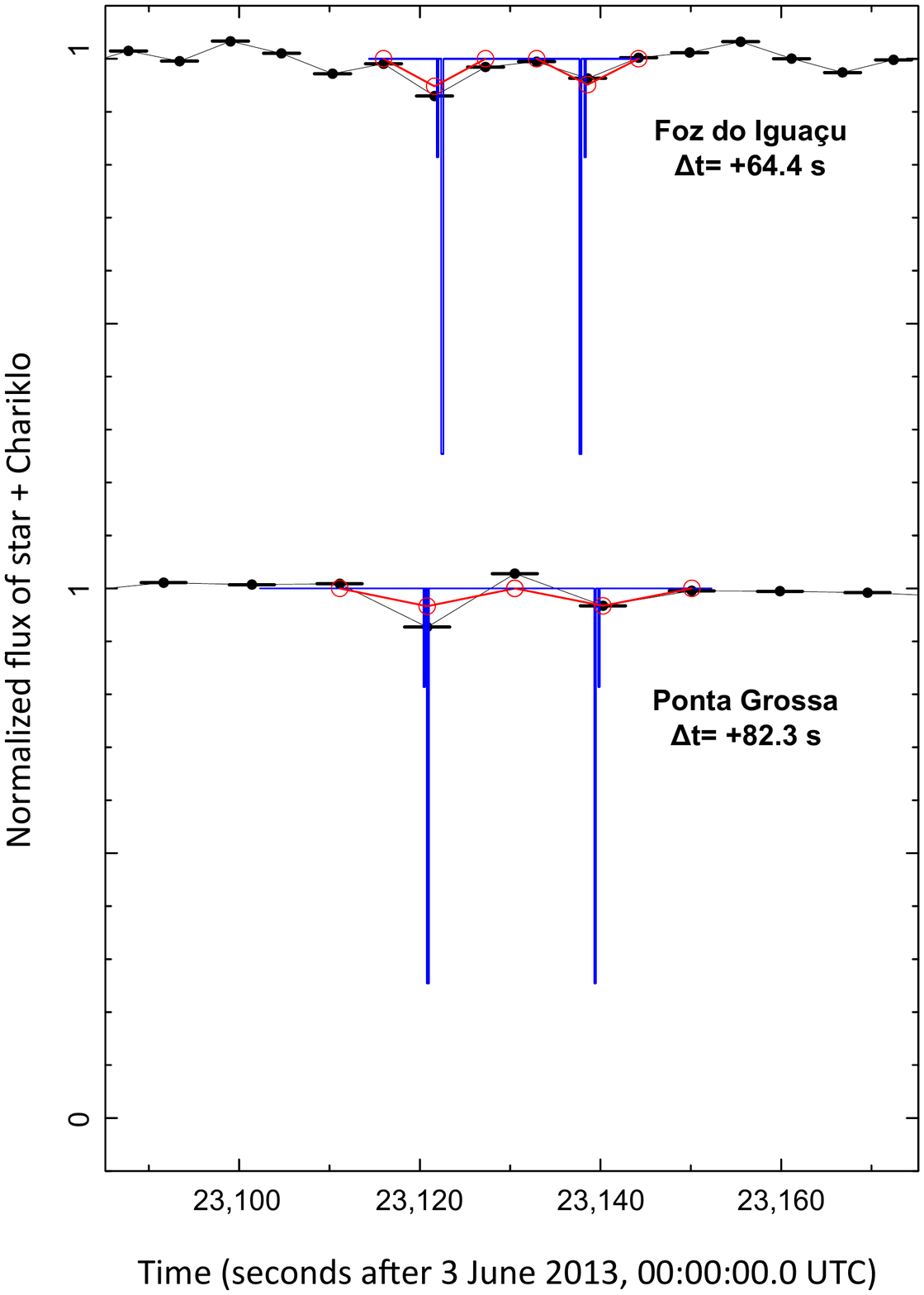} ~\ 
\includegraphics[totalheight=9.2cm,trim=11mm 1.7cm 2.9cm 4.1cm, clip]{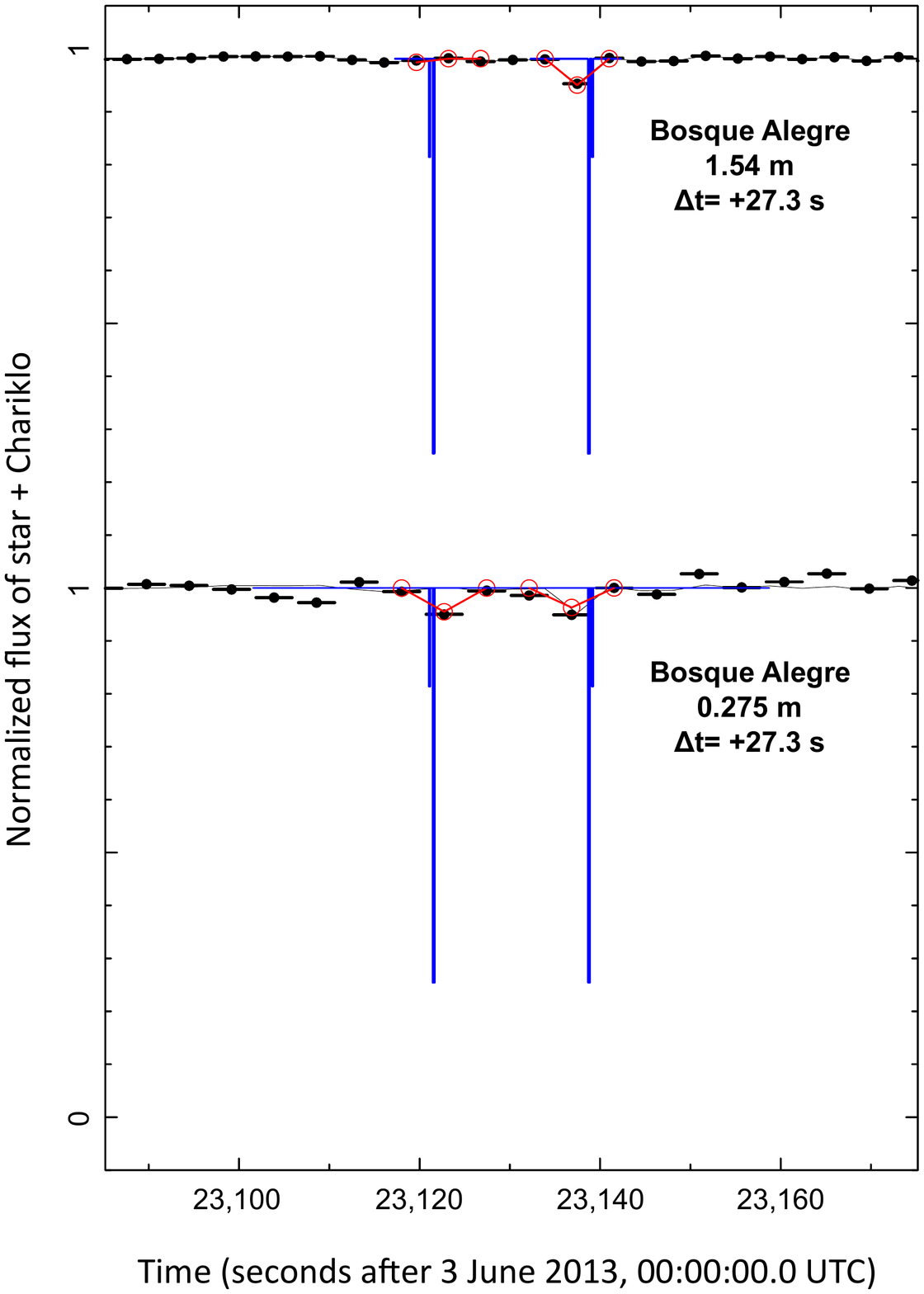}
}
\caption{ 
\vspace{-0.3cm}
\normalsize
\sf
\textbf{Extended Data Figure \ref{fig_all_fits} $|$ The fits to all the ring and Chariklo events.} 
The black dots are the data points and 
the horizontal bars indicate the corresponding intervals of acquisition.
The light curves are normalized between zero and unity (the latter corresponding to the 
full flux from the star plus Chariklo), and are shifted vertically for better viewing.
They are also displaced in time by the indicated amount $\Delta t$, 
in order to align the middles of the ring events.
The blue curves represent the ring model used to generate the synthetic  
profiles, plotted in red.
The ring widths and optical depths for La Silla (Danish and TRAPPIST telescopes)  
are taken from the Extended Data Table~\ref{tab_ring_danish}.
For all the other fits, we have used a unique ring model defined 
by the  ring geometry described in ED Table~\ref{tab_ring_geo}, 
and by the following widths and optical depths:
$W_{\sf C1R}= 6.6$~km,        $\tau_{\sf N,C1R}= 0.38$,
$W_{\sf C2R}= 3.4$~km and $\tau_{\sf N,C2R}= 0.06$,see footnotes of ED Table~\ref{tab_ring_timings}.
An expanded view of the fits to the Danish data is provided in Fig.~\ref{fig_fit_danish}.  } \label{fig_all_fits}
\end{figure}


\setlength{\tabcolsep}{1mm}
\renewcommand{\arraystretch}{1.1}  
\begin{table}
\vspace{-20mm}
\begin{center}
\hspace{-2.2cm} 
\footnotesize
\caption{  
\sf 
\textbf{Extended Data Table~\ref{tab_obs} $|$ Circumstances of observations. 
\label{tab_obs}}
}
\begin{tabular}{lllll}
\hline
Site   & Coordinates    & Telescope   &  Detector   & Observers    \\
        & Lat dd:mm:ss   & Name &  Camera &    \\
       & Lon dd:mm:ss  & Aperture & Exposure (s) &    \\
       & Altitude (m)         & Filter &   Cycle time (s)  &  Notes   \\
\hline
\hline
\multicolumn{5}{c}{Main body and ring positive detections.} \\
\hline
La Silla &  29$\dg$ 15' 21.3'' S   & Danish  & Lucky Imager  & C. Snodgrass,  \\
- Chile    & 70$\dg$ 44' 20.2''  W  & 1.54~m &  0.1     & J. Skottfelt, M. Rabus,  \\
           &  2336 & Z &  0.10091 & U. G. Jorgensen   \\
\hline
La Silla & 29$\dg$ 15' 16.6'' S    & TRAPPIST  & FLI PL3041-BB     & E. Jehin      \\
 - Chile     & 70$\dg$  44' 21.8''  W  & 0.6~m  & 4.5  & C. Opitom    \\
           & 2315     & no filter   &  5.7976456~s & Some cycles 6.15-6.19~s \\
\hline
Cerro Tololo   & 30$\dg$ 10' 03.4'' S      & PROMPT & U47-MB  &   J. Pollock  \\
- Chile           & 70$\dg$ 48' 19.0''  W    & 0.4~m$^{(\ast)}$  & 0.7   &    \\
                   & 2207 &  no filter    & 2.0  & Some 4~s cycles. \\
\hline
Cerro Pach\'on &  30$\dg$ 14'  16.8'' S     &  SOAR &  SOI   &   R. Duffard \\
   - Chile         &  70$\dg$ 44'  01.35'' W  & 4~m &  3 &   \\
                      & 2738 &  R     & 3.6--4.6   & Unstable cycling. \\
\hline
\hline
\multicolumn{5}{c}{Ring positive detections.} \\
\hline
Santa Martina$^{(\dag)}$  & 33$\dg$ 16' 09.0'' S  & M16  & Raptor/Merlin127  & R. Leiva Espinoza   \\
   -   Chile        & 70$\dg$ 32' 04.0'' W  & 0.4~m   &  1.0   & L. Vanzi   \\
                      & 1450       & no filter  &   1.0016887  &     \\
\hline
Bosque Alegre  & 31$\dg$ 35' 54.0'' S  & EABA  & Apogee AU9     &   C. Colazo    \\
 - Argentina      & 64$\dg$  32' 58.7''  W  & 1.54~m  & 3  & E.M. Schneiter  \\
                       & 1250    & no filter &  3.5616717 &  R. Melia \\
\hline
Bosque Alegre  & 31$\dg$ 35' 54.0'' S  & ORBA C11& Apogee AU10   &   C. Colazo    \\
 - Argentina     & 64$\dg$  32' 58.7''  W  & 0.275~m  & 4 &   E.M. Schneiter   \\
                       & 1250    & no filter &  4.71 &   R. Melia  \\
 \hline
Ponta Grossa    & 25$\dg$ 05' 22.2'' S    & RCX 400 &  SBIG STL6E     &  M. Emilio     \\
 - Brazil              & 50$\dg$  05'' 56.4'  W  &  0.4~m   & 5 & L. Mehret  \\
                        & 909     & no filter    & 9.78 &    \\
\hline
Foz do Iguaçu  & 25$\dg$ 26' 05.4''  S    & C11  &  SBIG ST-7XME  & D. I. Machado     \\
  - Brazil              & 54$\dg$ 35'  37.4''  W  &  0.275~m   &  4  & L. Lorenzini  \\
                      & 185    &  no filter   &  5.65 &    \\
\hline
Cerro Burek  & 31$\dg$ 47' 12'' S    & ASH  & SBIG-STL11000   &   N. Morales   \\
  - Argentina              & 69$\dg$ 18' 25''  W  &  0.45~m   &  7  & The ring event was  \\
                      & 2665   &  no filter   &  8.4 &   marginally detected.   \\
\hline
\hline
\multicolumn{5}{c}{Negative detections.} \\
\hline
Santa Rosa  & $ 36\dg$ 38' 16'' S    & El Catalejo  & Meade DSI-I   & J. Spagnotto     \\
  - Argentina              & 64$\dg$ 19' 28''  W  &  0.20~m   &  5  &   \\
                      & 182    &  no filter   &  1 &    \\
\hline 
San Pedro  & 22$\dg$ 57' 12''S    & Planewave & Apogee U42  & A. Muray     \\
~de Atacama  & 68$\dg$ 10' 48''  W  &  0.50~m   &  2  &  F. Char \\
 - Chile$^{(\P)}$                     & 2400   &  no filter   &  2.871 & B. Sandness   \\
\hline
Montevideo  & 34\dg 45' 20'' S    &  OALM  & FLI PL0900   &   S. Roland   \\
- Uruguay      & 56\dg 11' 23'' W  &  0.46~m   &  2 &  N. Martinez \\
                     & 130    &  no filter   &  2.83 & passing clouds.   \\
\hline
\hline 
\multicolumn{5}{c}{The following stations were clouded out during the event.}\\
\hline
\multicolumn{5}{l}{ Aigua-OAGA / Uruguay;  Buenos Aires / Argentina} \\
\multicolumn{5}{l}{ Itajub\'a-OPD; Rio de Janeiro-ON / Brazil.}\\
\hline
\end{tabular} 
\end{center}
$^{(\ast)}$ Three identical telescopes were used (P1, P3, P5), with exposures starting 0.7~s from each other. \\
$^{(\dag)}$ Universidad Catolica Observatory of Santa Martina (UCOSM).\\
$^{(\P)}$ The ASH2 (0.4~m) and ADO (0.38~m) telescopes were also used in this site, with larger cycle times.
\end{table}


\renewcommand{\arraystretch}{1.5} 
\begin{table}[!h]
\begin{center}
\sf
\footnotesize
\begin{tabular}{llllll}
\hline
\hline
Site  & N$^{(\dag)}$ &  \multicolumn{1}{c}{$t_{\sf C1R}^{(\ast)}$} & $\chi^2_{\sf dof}$ & \multicolumn{1}{c}{$t_{\sf C1R}^{(\ast)}$} & $\chi^2_{\sf dof}$ \\
\hline
 & & \multicolumn{2}{c}{ingress}  & \multicolumn{2}{c}{egress}  \\
\hline
Santa Martina  &
5 &
23,121.03$\pm$0.29 &
1.48  &
23,131.811$\pm$0.025 &
4.61 \\
Bosque Alegre/1.54m  &
3 &
(23,094.265$\pm$0.17)$^{(\S)}$ &
1.08 &
23,111.44$\pm$0.14 &
0.27 \\
Bosque Alegre/0.275m  &
3 &
(23,095.45$\pm$1.85)$^{(\star)}$ &
0.39 &
(23,109.45$\pm$1.75)$^{(\star)}$ &
0.94 \\
SOAR  &
3 &
(23,118.8$\pm$1.3)$^{(\ddag)}$  &
1.47  &
(23,138.4$\pm$1.4)$^{(\ddag)}$ &
0.11 \\
PROMPT  &
11 &
23,120.046$\pm$0.011 &
1.06 &
(23,140.445$\pm$0.155)$^{(\S)}$ &
0.81 \\
TRAPPIST  &
3 &
23,120.9$\pm$1.9 &
2.99  &
23,140.88$\pm$0.53 &
2.01  \\
Ponta Grossa  &
3 &
23,038.6$\pm$2.5 &
6.70 &
23,058.0$\pm$2.5 &
1.93 \\
Foz do Igua\c{c}u  &
3 &
23,057.5$\pm$1.7 &
1.64 &
23,074.1$\pm$2.0 &
0.42 \\
\hline
\hline
\end{tabular}
\end{center}
\caption{  
\sf
\textbf{Extended Data Table~\ref{tab_ring_timings} $|$
Timings of the ring events besides the Danish telescope.}%
\label{tab_ring_timings}}
%
$^{(\ast)}$Assuming that rings C1R and C2R have widths and optical depths
that are averages of the ingress and egress values at the Danish telescopes; i.e. 
$W_{\sf C1R}= 6.6$~km,        $\tau_{\sf N,C1R}= 0.38$,
$W_{\sf C2R}= 3.4$~km and $\tau_{\sf N,C2R}= 0.06$, 
radial separation between rings C1R and C2R: 14.2~km. The fitted time for the middle of ring C1R, seconds after 00:00:00 UT (3 June 2013). \\
$^{(\dag)}$The number of fitted data points. \\
$^{(\S)}$Obtained from non-detections of the rings, see text. \\
$^{(\star)}$Confirms the ring detection at Bosque Alegre, but is not  used in the
fit of the ring orbits due to larger error bars compared to the 1.54~m telescope. \\
$^{(\ddag)}$The SOAR timings are probably affected by a systematic offset, and are not used in the
fits of the ring orbits and Chariklo's limb shape, see text.\\
As Cerro Burek chord does not bring further constraints to our analysis, it is not listed here.
\end{table}


\setlength{\tabcolsep}{3mm}
\renewcommand{\arraystretch}{1.2} 
\begin{table}[!h]
\begin{center}
\sf
\footnotesize
\begin{tabular}{lllllll}
\hline
\hline
Ring   & \multicolumn{1}{c}{$N$} 
           &  \multicolumn{1}{c}{$\chi^2_{\sf dof}$}
           & \multicolumn{1}{c}{$t$} 
           &  \multicolumn{1}{c}{$W$ (km)}
           &  \multicolumn{1}{c}{$\tau_{_N}$} 
          &  \multicolumn{1}{c}{$E_\tau$ (km)}  \\
\hline
\multicolumn{7}{c}{ingress} \\
\hline
C1R &
10 & 
0.95 & 
23,121.168$\pm$0.0007 &
$6.16\pm0.11$ & 
$0.449\pm0.009$ & 
$2.77\pm0.04$ \\
C2R  & 
8  & 
0.69 & 
23,120.765$\pm$0.011 &
$3.6^{+1.3}_{-2.0}$ & 
$0.05^{+0.06}_{-0.01}$ & 
$0.18\pm0.03$  \\
\hline
\multicolumn{5}{l}{Radial separation (C2R minus C1R): 14.6$\pm$0.4 km} &  &  \\
 \multicolumn{4}{l}{Gap between rings C1R and C2R: }    &  9.0$\pm$0.4 km & $\tau_N <$ 0.004 & $E_\tau <$0.04~km      \\
\hline
\hline
\multicolumn{7}{c}{egress} \\
\hline
C1R &
10 & 
1.19 & 
23,140.462$\pm$0.0012 &
$7.17\pm0.14$ & 
$0.317\pm0.008$ & 
$2.28\pm0.03$ \\
C2R  & 
8   & 
0.71 &
23,140.847$\pm$0.006 &
$3.4^{+1.1}_{-1.4}$ & 
$0.07^{+0.05}_{-0.03}$ & 
$0.24\pm0.02$ \\
\hline
\multicolumn{5}{l}{Radial separation  (C2R minus C1R): 14.1$\pm$0.2 km} &  & \\
 \multicolumn{4}{l}{Gap between rings C1R and C2R: }    &  8.3$\pm$0.2 km & $\tau_N <$ 0.004 & $E_\tau <$0.05~km   \\
\hline
\hline
\end{tabular}
\caption{ 
\normalsize
\sf
\textbf{Extended Data Table~\ref{tab_ring_danish} $|$
Physical parameters of rings C1R and C2R.}
\label{tab_ring_danish} %
\sf
$N$: number of fitted data points
(there are $M=3$ free parameters: $t$, $W$ and $\tau_{_N}$);
$\chi^2_{\sf dof}= \chi^2/(N-M) $, the $\chi^2$ per degree of freedom; 
$t$: mid-times of ring events, in seconds after 00:00:00 UT (3 June 2013). 
The error bars quoted here are internal to the fits, and are given at 1$\sigma$ level.
The  error bars in absolute time, $\pm$0.014 seconds, are larger than the error
bars on the relative times reported here, see text; 
$W$: radial width, measured in the plane of the rings 
(using the ring pole given in Table~\ref{tab_ring_prop}); 
%
%
$\tau_{_N}$: normal optical depth; 
%
$E_\tau= W \cdot \tau_{_N}$: equivalent depth.
}
\end{center}
\end{table}


\renewcommand{\arraystretch}{1.5} 
\begin{table}[!h]
\begin{center}
\sf
\footnotesize
\begin{tabular}{llllll}
\hline
\hline
\multicolumn{5}{c}{Elliptical fit to the ring events, projected in the plane of the sky} \\ 
\multicolumn{5}{c}{Number of ring positions fitted: 14, number of adjusted parameters: 5, $\chi^2_{\sf dof}= 1.48$} \\ 
\hline
$f_{\sf c}$ (km) & 
$g_{\sf c}$ (km) & 
$a$ (km) & 
$e$ & 
$B$ (deg)  &
\multicolumn{1}{c}{$P$ (deg)} \\
\hline
-2734.7$\pm$0.5    &    +793.8$\pm$1.4    &    390.6$\pm$3.3    &    0.444$\pm$0.006    &   
+33.77$\pm$0.41 &  \multicolumn{1}{c}{\textbf{-61.54$\pm$0.14}} \\
 & & & & & or 118.46$\pm$0.14 \\
\hline
\hline
 & \multicolumn{3}{l}{Ring pole (equatorial J2000)}  \\ 
\hline
\hline
   &  \multicolumn{1}{l}{$\alpha_{\sf p}$ (deg )} & 
\multicolumn{3}{l}{$\delta_{\sf p}$  (deg)}    \\
\hline
 &  \multicolumn{1}{l}{\textbf{151.30$\pm$0.49}} &
{\textbf{41.48$\pm$0.21}} & \multicolumn{3}{l}{\textbf{Solution 1 (preferred, see text)}} \\
\hline
  &   \multicolumn{1}{l}{26.96$\pm$0.29} & 
03.44$\pm$0.31 &   \multicolumn{3}{l}{Solution 2} \\
\hline
\hline
\end{tabular}
\end{center}
\caption{  
\sf
\normalsize
\textbf{Extended Data Table~\ref{tab_ring_geo} $|$
Ring geometry.}
\label{tab_ring_geo} 
The elliptical fit uses the timings and associated error bars 
of Extended Data Table~\ref{tab_ring_timings}, and the methodology 
used in our previous works\cit{Sicardy11,BragaRibas13}.
Definitions of listed parameters are (see also text): 
$(f_{\sf c},g_{\sf c})$: centre of the ring in the plane of the sky, 
counted positively toward local celestial east and north, respectively.
This is the offset to apply to Chariklo's ephemeris in order to fit the observations.
Transformation in arcsec can be performed using a geocentric distance 
$D=2.031 \times 10^9$~km during the occultation.  
We use here the star position given in Eq.~\ref{eq-star}
and the JPL\#20 Chariklo ephemeris; 
$a$: apparent semi-major axis of the ring projected in the plane of the sky;
$e$: aspect ratio of the ring $e= (a-b)/a$; 
where $b$ is the apparent semi-minor axis of the ring projected in the plane of the sky;
$P$: The position angle defines the angle between the semi-minor axis of the ring projected in the plane of the sky, counted positively from celestial north to celestial east. By convention, it refers to the projected semi-major axis that corresponds to superior conjunction.
$B$: the ring opening angle, calculated from $|\sin(B)|= 1 - e$. 
Equivalently, it is the absolute value of the elevation of the observer above the ring plane. }
\end{table}


\renewcommand{\arraystretch}{1.5} 
\begin{table}[!h]
\begin{center}
\sf
\footnotesize
\begin{tabular}{lllllll}
\hline
\hline
Site  & N &  \multicolumn{1}{c}{ingress} & $\chi^2_{\sf dof}$ & \multicolumn{1}{c}{egress} & N & $\chi^2_{\sf dof}$ \\
\hline
Danish$^{(\ast)}$  & 10 & 23,127.861$\pm$0.014    & 0.30  & 23,133.188$\pm$0.014 & 10 & 3.3  \\
TRAPPIST        & 9 & 23,127.893$\pm$0.019       & 0.84  & 23,133.155$\pm$0.007 & 10 & 1.47 \\
PROMPT          & 22 & 23,124.835$\pm$0.009 & 0.71 & 23,135.402$\pm$0.015 & 26 & 0.58 \\
SOAR$^{(\dag)}$  & 3   & (23,124.34$\pm$0.59)          & 0.72  & (23,134.597$\pm$0.009) & 3 & 0.57 \\
\hline
\hline
\end{tabular}
\caption{  
\textbf{Extended Data Table~\ref{tab_Chariklo_timings} $|$ Timings of the Chariklo event.
}
\label{tab_Chariklo_timings}}
\end{center}
\sf 
$^{(\ast)}$ These timings were corrected by -1.622 seconds using the TRAPPIST times, see text for details.\\
$^{(\dag)}$ These timings may be affected by a delay of about 0.5~s, and are not used for Charilklo's limb
fitting, see text for details.
\end{table}


\renewcommand{\arraystretch}{1.5} 
\begin{table}[!h]
\begin{center}
\sf
\footnotesize
\begin{tabular}{llllll}
\hline
\hline
$~~f_{\sf c}$ (km)$^{(\ast)}$ & 
$~~g_{\sf c}$ (km)$^{(\ast)}$ & 
$R_{eq}$ (km) & 
$R_{\sf equiv} \sf{(km)}^{(\dag)}$ &  
$~~~~~~e$   &
$~P$ (deg)$^{(\ast)}$ \\
\hline
-2734.7$\pm$0.5  &  
+793.8$\pm$1.4  &  
144.9$\pm$0.2  &  
126.9$\pm$0.2 & 
0.213$\pm$0.002 &  
-61.54$\pm$0.14   \\
\hline
\hline
\multicolumn{6}{c}{Geocentric position derived from the occultation (J2000)$^{(\S)}$}  \\
\hline
\multicolumn{2}{c}{ Time (UT)} & \multicolumn{2}{l}{Right Ascension} & Declination \\
\hline
\multicolumn{2}{c}{06:30:00} &  \multicolumn{2}{l}{ $\sf{ 16^{\sf h}~56^{\sf min}~06.4618^{\sf s}\pm0.006^{\sf s}}$ }  &  \multicolumn{2}{l}{$-40\dg 31'~30.009'' \pm 0.002''$}  \\
\hline
\hline
\end{tabular}
\caption{ 
\textbf{Extended Data Table~\ref{tableChar} $|$
Chariklo Physical Properties.}
\label{tableChar} }%
\end{center}
\sf 
$^{(\ast)}$ The centre and position angle of the elliptical fit to the Chariklo chords, were taken from the ring fit, see Extended Data Table \ref{tab_ring_geo}. \\
$^{(\dag)}$ $R_{\sf equiv}= R_{eq} \sqrt{1-\epsilon}$ is the radius of a circle having the same apparent area of the fitted ellipse. \\
$^{(\S)}$ The error is largely dominated by the star position error determination, not by the limb fit.
\end{table}

\sf 
\setlength{\textwidth}{16cm}            
\setlength{\parindent}{0mm}    


\thispagestyle{empty}
\setcounter{page}{1}

\huge
\noindent
\color{darkblue}{\sc Supplementary Information} \cp\\
\normalsize 
\hl

\section{\sf Prediction and observations}

The star UCAC4 248-108672  was found as a potential candidate for an occultation by the Centaur object  (10199) Chariklo in 3 June 2013.
This event was discovered in a systematic search\cit{Camargo13} conducted at the Max-Planck 2.2~m telescope 
of the European Southern Observatory (ESO). 
The sky paths of 39 Transneptunian and Centaur objects from 2012.5 to 2014 were imaged, 
and stars up to magnitude $R \sim$ 19 were astrometrically measured against the UCAC4 frame\cit{Zacharias13}. 

The 3 June 2013 event involved one of the brightest stars  to be occulted by Chariklo in 2013, 
with $R\sim$ 12.4 and $K\sim$ 7.4. 
The shadow path was predicted to cross Brazil, Chile, Argentina and Uruguay on 
3 June 2013 around 06:25~UT (Extended Data Figure \ref{fig_map}). 
Astrometric updates of the star and Chariklo were performed at the beginning of May 2013
at the ESO 2.2~m telescope and 
at the 0.6~m Boller \& Chivens telescope of Pico dos Dias Observatory (OPD), Brazil.
These observations provided the following ICRF (J2000) star position:
\begin{equation}
\left\{%
\begin{array}{l}
\sf \alpha=16^{\sf h} ~56^ {\sf min} ~06.4876^ {\sf s} \pm 0.0058^ {\sf s} \\ 
\sf \delta= -40^\circ ~31' ~30.205'' \pm 0.0023'' \\
\end{array}
\right.
\label{eq-star}
\end{equation}
Further observations with the large field of view of the Astrograph for the Southern Hemisphere (ASH)
0.45m telescope, Cerro Burek, Argentina, confirmed the predictions above, 
as Chariklo and the candidate star could be imaged in the same field of view. 

An observational campaign was organized with 14 different sites involved (Extended Data Table~\ref{tab_obs}). 
Information about the event, finding charts and tips were made available in a dedicated web-page 
(\url{http://devel2.linea.gov.br/~braga.ribas/campaigns/2013-06-03_Chariklo.html}). 

\section{\sf Timing issues}

Image calibration and light curve derivations proceeded as for our previous observations\cit{Sicardy11,Ortiz12,BragaRibas13}.
Specific information, about some timing issues on some instruments, are given below.

\textit{Danish telescope} - 
The best data in terms of signal-to-noise ratio (SNR) per second of integration, were obtained at the 1.54~m Danish telescope
at La Silla, a site of the European Southern Observatory (ESO) in Chile.
The camera, a \emph{Lucky Imager} from the 
Microlensing Network for the Detection of Small Terrestrial Exoplanets 
(MiNDSTEp) consortium, acquired 6,000 images with exposure time 
0.1~s and cycle time 0.10091~s in Z filter (a long-pass filter with cut-off below 650 nm and 
limited in the red by the sensitivity of the CCD chip), 
starting at  06:20:07 UT, see ED Table~\ref{tab_obs} for more details.
The images were taken by cubes of 100, each separated by about 0.3~s required for readout and saving.
None of the ring events, and star disappearance and re-appearance from behind Chariklo occurred
during those blind time intervals.
%
%

The \emph{Lucky Imager} uses GPS (Global Positional System) to synchronize local dual NTP (Network Time Protocol) servers to provide a high precision time reference.
However, comparison with the timestamps from the TRAPPIST\cit{Jehin11} (TRAnsiting Planets and PlanetesImals Small Telescope) at the same observatory shows 
that the Danish timing was ahead by about 1.6~seconds. 
This appears to be introduced by the software controlling the camera.
It has a high internal  relative precision over the course of an observing sequence and writes timestamps 
to the output file every 100 frames (10 seconds), but contains an offset from the absolute reference time 
which varies with observing sequences taken on different nights. 
The origin of this offset remains unknown. 
This said, the TRAPPIST timing is preferred because it has been used in various occultation campaigns 
in the previous years without showing evidence of timing offsets relative to other stations. 
In particular, a stellar occultation by Pluto observed on 4 May 2013 from six stations in Argentina, Brazil and
Chile (including TRAPPIST and the Danish telescope) showed that all the stations had consistent timings,
except for the Danish, whose timing was lagging the other stations by about 2.5~seconds.
So, we decided to use the Chariklo occultation timings obtained at TRAPPIST, 
and then align the mid-occultation times obtained at the Danish telescope with TRAPPIST, 
which requires a correction of -1.622~seconds to the Danish timings.
All the Danish timings quoted in the letter and here are given with this correction applied.
Note that the uncertainty in the absolute timing of the Danish data comes
from the uncertainties of the TRAPPIST occultation fit, or $\pm$ 0.014~s.
However, the relative timings of the Chariklo and ring occultations obtained 
at the Danish have a better accuracy of the order of the millisecond (see ED Table~\ref{tab_ring_danish}).

\textit{SOAR} - 
The Southern Astrophysical Research (SOAR) 4.1~m Telescope at Cerro Pach\'on, Chile,  acquired images in R band, with 3~s exposure time and 4.5~s average cycle time.
The timing of the star re-appearance from behind Chariklo at SOAR indicates a discrepancy of about 0.5~s
when compared to the same event observed at PROMPT (at CTIO),  which is located $\sim$~10~km 
away from SOAR.
%
%
We have no way to re-calibrate the SOAR timing using another nearby station as a reference.
The shift of about 0.5~s that needs to be applied to the SOAR chord to be compatible with the PROMPT timing 
can be explained by a shutter delay in the SOAR Optical Imager (SOI). 
Shutter delays of the order of the second are not unusual in cameras 
that are not specifically designed for very high timing accuracy applications.  
Large mechanical shutters of large format CCDs coupled to computer software and network issues 
can result in delays of the image acquisition compared to the time stamped in the fits headers of the images. 
Unfortunately we have not been able to calibrate this delay for SOI,
and the corresponding timings published in this work have not been corrected.
Consequenty, although the SOI data are important to confirm the Chariklo and the ring events,
its timings are not used for the ring orbit fitting, nor for the Chariklo limb fitting.

\section{\sf Fits to the ring events\label{RingFit}}

The Danish light curve (Fig.~\ref{fig_DanishCurve}) clearly shows the main body occultation surrounded  
by two prominent  drops (named C1R) which are themselves flanked by two smaller drops (C2R). 
In both cases, the flux between the drops C1R and C2R goes back to unity, 
indicating that no material is detected between the two events, see the discussion below (Sec. \ref{fitDanish}).

\subsection{\sf Diffracting ring model}

The rings are modelled as sharp-edged, semi-transparent bands of transmission $T$
(along the line of sight)  and 
apparent width $W_{\sf app}$ (along the occultation chord, see Fig.~\ref{fig_rings}).
We use an algorithm\cit{Roques87} that describes the Fresnel diffraction 
on such bands, with further convolution 
by 
the finite bandwidth of the instrument, 
the stellar diameter  
and the finite integration time.

To generate the ring profile, we actually need $v_\perp$, the perpendicular velocity of the star relative to the local ring projected in the plane of the sky. 
This requires the knowledge of the ring geometry,
i.e. its opening angle $B$ and position angle $P$, see ED Table~\ref{tab_ring_geo}. 
These two angles are determined by an elliptical fit to the various events (the green segments in Fig.~\ref{fig_rings}).

The optical depth along the line of sight, $\tau$, is related to the transmission by $T=\exp(-2\tau)$.
The factor two is caused by the diffraction on individual particles\cit{Cuzzi85}, 
that scatters the light over an Airy scale $F_A \sim (\lambda/2r)D$. 
Here 
$\lambda$ is the wavelength of observation,
$r$ is the radius of the particles,  and
$D$ is the distance to Chariklo. 
Assuming that $r$ is of the order of meters at most, 
as in Saturn's\cit{Cuzzi09} or Uranus' rings\cit{Esposito91},
and using $D = 2.031 \times 10^{9}$~km, 
we obtain  $F_A \go 50$~km in the red band, significantly larger than $W_\perp$, the local perpendicular width of the ring in the plane of the sky, that is of the order of only 5~km
(ED Table~\ref{tab_ring_danish}).

Assuming a multi-layer particle ring, $\tau$ is related to the normal optical depth $\tau_{_N}$ by $\tau_{_N}=\tau\cdot|\sin(B)|$. 
Finally, the radial width $W$ of the ring (in the ring plane) is derived from $W_\perp$
using the pole position given in ED Table~\ref{tab_ring_geo}.
For a multi-layer ring, a physically relevant quantity is the equivalent depth $E_\tau= W \cdot \tau_{_N}$, which is related
to the total amount of material contained in the ring at the sub-occultation point\cit{French91}. The star radius $r_{\sf star}$ projected at Chariklo, about 1~km (see below), 
is comparable to the Fresnel diffraction scale, $F_F= \sqrt{\lambda D/2} \sim 0.8$~km in the red band. 
Hence it has to be correctly evaluated for a correct fit to the ring events.

To estimate the angular diameter of the occulted star, we consider its following magnitudes 
$m_B = 15.24 \pm 0.11$, 
$m_V=13.38 \pm 0.07$ (UCAC4\cit{Zacharias13}), 
$m_J=8.67 \pm 0.04$, 
$m_H = 7.62 \pm 0.04$, 
$m_K=7.26 \pm 0.08$ (2MASS\cit{Skrutskie06}),  
$m_{3.35\,\mu m} = 6.98 \pm 0.05$ and
$m_{4.60\,\mu m} = 7.14 \pm 0.05$ (WISE\cit{Cutri12}). 
The comparison of the observed spectrum of the star (ED Fig.~\ref{StarSpec}) with models\cit{CastelliKurucz03} of different temperatures indicates an effective temperature between 3,500 and 4,000 K for the occulted star. For the star's spectral energy distribution, we therefore choose the $T_\mathrm{eff} =3,750$\,K template spectral energy distribution ($\sf \log(g)=2.0$, solar metallicity). For the reddening, we adopt $R_V = A_V / E (B-V) = 3.1$, a standard interstellar dust model\cit{Fitzpatrick99}. A simultaneous fit of the photospheric angular diameter of the star and the selective extinction $E(B-V)$ to the available photometry gives a relatively high extinction $E(B-V) = 0.68 \pm 0.10$ and an apparent angular radius of $0.111 \pm 0.002$\,milliarcseconds. The derived extinction is typical of a star located close to the Galactic plane (galactic latitude $+1.7^\circ$), that could be a red giant at a distance of the order of 1\,kpc ($M \approx 0$). The resulting angular diameter corresponds to a star radius projected at Chariklo distance of $r_\mathrm{star} = 1.09 \pm 0.02$\,km.


\subsection{\sf Fits to the Danish ring events\label{fitDanish}}

Due to its higher acquisition rate (10 Hz),  the Danish telescope
is the only  instrument that could resolve the rings C1R and C2R.   
Extended Data Table~\ref{tab_ring_danish} provides the three adjusted parameters of each fit:
the time $t$ of the mid ring event, 
the radial width $W$ (in the plane of the ring) and 
its normal optical depth $\tau_{_N}$
(providing the equivalent width $E_\tau= W \cdot \tau_{_N}$).
Also listed are the radial distances from ring C1R to ring C2R.
The fits are shown and discussed in Fig.~\ref{fig_fit_danish}. 
The error bars quoted in ED Table~\ref{tab_ring_danish} account for the error
bars stemming from the noise in the data when performing the diffracting fits, and
the errors bars caused by the uncertainty in the ring opening angle $B=33.77$$\pm$0.41~degrees (see ED Table~\ref{tab_ring_geo}),  which modify the projections of the various quantities of interest from the sky plane to the ring plane.
Although being of the same order of magnitude, the error bars caused by the noise in the data 
always dominate the error bars caused by the uncertainty in $B$.

No material is detected between the events C1R and C2R,
to within the noise level (Fig.~\ref{fig_fit_danish}). Considering that the signal to noise
ratio of the Danish light curve is 64 per data point (Fig.~\ref{fig_DanishCurve}),
the 1$\sigma$ upper limit  of normal optical depth $\tau_{N,\sf gap}$ for inter-ring material is
given by $\exp(-2\tau_{N,\sf gap}/|\sin (B) |)= 1 - (1/64)$, or $\tau_{N,\sf gap} < 0.004$.
We also provide in ED Table~\ref{tab_ring_danish} the corresponding upper limits
for the equivalent depth of that material.

In the rest of the text, the term ``ingress", resp. ``egress", will hereby refer to the first, 
resp. second,  of a pair of ring events at a given station.

\subsection{\sf Fits to the other ring events}

Besides those of the Danish telescope, none of the ring events are resolved.
They  are detected in one data point, with a drop of signal that only depends 
on the total equivalent depth of rings C1R and C2R.
To determine the timings of the events, we have fitted a model that includes both rings,
assuming that their widths, optical depths and radial separation are constant in longitude.
The values of those parameters are given in ED Table~\ref{tab_ring_timings}.

The fits to all the events are displayed in ED Fig.~\ref{fig_all_fits},
while the timings and the quality factor $\chi^2_{\sf dof}$ of the fits are listed in ED Table~\ref{tab_ring_timings}.
Note that the PROMPT ingress ring event is detected over two data points, 
but  they were in fact obtained at two separate telescopes with overlapping integration intervals 
(ED Table~\ref{tab_obs}).

The worst fit is obtained for the ingress Ponta Grossa ring event. 
A closer inspection of the fit (ED Fig.~\ref{fig_all_fits}, third panel) 
shows that part of the high value of the residual is due to the 
fact that the third fitted data point is higher than the synthetic point,
while the second data point, caused by the ring, 
is below the synthetic point by about the same amount (at 2.8$\sigma$ level). 
This shows that the observed discrepancy is likely to be caused by local noise. 
%
%
%
As mentioned in the letter, the fits to the C1R events at the Danish
reveals a small but significant variation between ingress and egress.
No significant variations of the equivalent depth of the ring system (C1R plus C2R)
with longitude are observed in the other events.

We also note that the rings are \textit{not} detected at ingress in the Bosque Alegre 1.54~m telescope data.
This is consistent  with an overhead interval of 0.56~seconds between two 
images. 
The ingress and egress events were detected with Bosque Alegre 0.275~m telescope, 
confirming that the ring is actually present in this region. 
As they present larger error bars than the \textit{non} detection from the 1.54~m telescope at the same site, 
they are not used for the ring fit.
The egress ring event at PROMPT was not detected either, with only a possible, very marginal
detection of ring C2R. 
This is again consistent with a blind interval of 0.60~seconds in the acquisition, 
due to an incomplete time coverage of the three PROMPT telescopes.
Assuming that the ring is uninterrupted, these non-detections are still useful as 
they provide a ring position with good accuracy, and thus 
constrain further the ring geometry, see the red segments in Fig.~\ref{fig_rings}.

In total, we have fifteen ring detections in thirteen different ring positions 
(as in La Silla and Bosque Alegre more than one telescope was used, and we are counting here each of Danish double events C1R+C2R as a single event), 
as well as the two non-detections discussed just above. 
At Cerro Burek the ring detections were marginal (less than 2.4$\sigma$), 
due to the long integration time and limited signal-to-noise ratio, 
so they do not bring any further constraints to our ring analysis.
Note however, that the non-detection of Chariklo occultation itself may provide relevant upper limits
for the size of the main body.
Excluding the two SOAR events, because of timing problems, 
the rest of positive and negative detections provide the ring geometry shown in Fig.~\ref{fig_rings}.
An elliptical fit to all the positions (weighted according to their respective radial error bars) 
provides 
the centre $(f_{\sf c}, g_{\sf c})$ of the rings, 
the apparent semi-major axis $a$ of ring C1R (the semi-major axis of the ring C2R being 14.2~km larger), 
their opening angle $B$ and
position angle $P$, see ED Table~\ref{tab_ring_geo} for details.
Assuming that the rings have circular orbits 
(so that $a$ coincides with the orbital radius of ring C1R), 
their apparent ellipticity and orientation provide two possible ring poles, 
one being preferred as it explains Chariklo's photometric and spectroscopic variations,
see the letter.

\section{\sf Ring photometry}


The ring intrinsic brightness may be estimated by assuming that 
the dimming of Chariklo's system\cit{Belskaya10} between 1997 and 2008 
is entirely due to the gradual disappearance of the rings, 
as their opening angle $B$ went from 60 to 0 degrees, 
as predicted from our preferred ring orbital solution given in Table~\ref{tab_ring_geo}
(NB. the solution 2 predicts that the rings were observed edge-on in 1995, 
and wide-opened by about 50~degrees in 2008).
Chariklo's reflectivity $(I/F)_C(\alpha)$ at phase angle $\alpha$ is given by $(I/F)_C(\alpha)= p_C  \Phi_C(\alpha)$,
where $I$ is the intensity of the light reflected by the surface, 
$p_C$ is Chariklo's  geometric albedo and $\Phi_C(\alpha)$ is its phase function.
As absolute magnitudes will be considered here, we take $\alpha=0$, so that $\Phi_C(0)=1$. 
Using the same notation for the ring surface, but with subscript $r$, we obtain 
the ratio of the fluxes received from the rings and Chariklo:
$$
\frac{F_r}{F_C}=
\frac{(I/F)_r}{(I/F)_C} \cdot \frac{S_r}{S_C}.
$$
Here $S_C= \pi R^2_C$ is Chariklo's apparent surface,  
where $R_C$  is the equivalent radius of the body.
The apparent surface area of a ring of radius $a$ with radial width $W$
is given by $S_r= 2\pi \sin(B) a W$.

Chariklo's absolute visual magnitude varied\cit{Belskaya10} from about 6.8 to 7.4 between 1997 and 2008,
corresponding to a reduction of brightness by a factor of 1.75 with
error bars that we estimate as $\pm$~0.15.
Thus, $F_r/F_C ~\sim~0.75 \pm 0.15$ in 1997.
%
%
The equivalent radius and geometric albedo of Chariklo are estimated\cit{Fornasier13} to be
$R_C= 124\pm9$~km and $p_C=0.035\pm0.01$.
%
We consider here that most of the flux comes from the ring C1R, which has 
$a_{C1R} = 391$~km and  $W = 6.6$~km (ED Table~\ref{tab_ring_danish}), and 
$B\sim 60$~degress in 1997, so that:
%
%
\begin{equation}
\left(\frac{I}{F}\right)_r \sim 0.09 \pm 0.04.
\label{eq_IFr}
\end{equation}
We may compare this value with the reflectivity of Saturn's A ring\cit{Hedman13}, 
which has an optical depth comparable to that of ring C1R, $(I/F)_{\sf ring~A} \sim 0.3$.
Conversely, Uranus rings\cit{Karkoschka01} $\alpha$ and $\beta$, which also have optical depths comparable
to that of C1R, $(I/F)_{\alpha,\beta} \sim 0.05$.
Thus, in terms of reflectivity, Chariklo's rings appear about three times darker than Saturn's A ring,
twice as bright as Uranus' $\alpha$ and $\beta$ rings, and about three times brighter than
Chariklo's surface.
Finally, we see that a ring to Chariklo flux ratio $F_r/F_C\sim$~0.75 in 1997 (with $B\sim 60\dg$)
would imply a flux ratio $F_r/F_C\sim$~0.5 as of today ($B\sim34\dg$).

%
%
%
%

Note that part of the photometric variation observed between 1997 and 2008 may be due
to the changing viewing geometry of Chariklo itself. 
For instance, if we accept that Chariklo is an oblate spheroid with flattening 0.33
(derived from an apparent flattening 0.212 at opening angle $B \sim$~34~degrees in June 2013,
see Section \ref{CharikloShape}), 
then we calculate that its apparent surface area varied by a factor of 1.23. This would reduce the ring contribution 
relative to Chariklo from 0.75 (as estimated above) to about 0.50, 
which would in turn darken the ring by a factor of 0.5/0.75~$\sim$~0.65.
As Chariklo's shape is not known, however, this correcting factor cannot be assessed precisely at present time.

\section{\sf Ring dynamics}

The closest analogue to Chariklo's rings in terms of structure is the Uranus system, 
with nine dense and narrow rings\cit{French91}, 
some of them (in particular the $\alpha$ and $\beta$ rings) comparing in 
widths and optical depths to the ring C1R. 
In terms of composition and optical depth, the presence of water ice  in Chariklo's rings (see the letter) make them akin to Saturn's outer A ring\cit{Colwell09}.
In addition, changes in the Chariklo's spectrum from 1999 to 2008\cit{Guilbert09} indicate that the rings' spectrum includes a steep slope below 0.55~$\mu$m, also seen in Saturn's A ring\cit{Hedman13}, where it is attributed to some contaminant, possibly  an organic compound or nanometre-sized grains of iron and hematite.


We note also that an icy body like Chariklo, of radius $\sim 125$~km, has a mass of the order of 
$10^{19}$~kg, so that the mean motion at orbital radius $\sim 400$~km is $n \sim 10^{-4}$~s$^{-1}$.
This is comparable to the orbital mean motions of Uranus' rings and of the outer part
of Saturn's main rings.
A collisional disk tends to stabilize  near a Toomre parameter $Q= c_s n/\pi G \Sigma \sim 1$,
where $c_s$ is the velocity dispersion, $G$ is the constant of gravitation and 
$\Sigma$ is the surface density.
Thus, similar particle composition, optical depths and mean motion,
suggest dynamical parameters like velocity dispersion, viscosity and thickness of Chariklo's rings similar to those of their Uranian and Saturnian cousins.
 
Finally, Uranus' rings evolve in a rich environment of small satellites which
probably have a close link to their origin and confinement, providing constraints
on their formation\cit{Tiscareno13,Colwell92}, and pointing toward a possible satellite system
near Chariklo.
This said, considering that the ring diameter subtends 0.08~arcsec as seen 
from Earth, direct imaging of any satellite near the rings is very challenging.
Consequently, all the estimations given below remain speculative, 
and are intended to provide order of magnitude considerations on the rings and their hypothetical companion satellites. 

An unperturbed collisional ring of width $W$ spreads out on a time scale\cit{Goldreich82} $t_{\nu}$, where $\nu$ is the kinematic viscosity of the ring:
\begin{equation}
t_\nu \sim \frac{W^2}{\nu},
\label{eq_spread}
\end{equation}
A typical value of $\nu$ is $\sim n h^2$, where $h$ is the ring thickness.
Taking $W \sim 5$~km (ED Table~\ref{tab_ring_danish}), 
we derive  $t_\nu \sim 10^4/h^2$~years, where $h$ is expressed in metres.
Assuming $h \sim$ a few metres (typical of Saturn's A ring\cit{Colwell09}), 
we obtain $t_\nu \sim$ a few thousand years. 
Moreover, a ring with a range of particle sizes around a typical value $R$,
spreads due to Poynting-Robertson (PR) differential drag, on a time scale\cit{Goldreich79}
\begin{equation}
t_{PR} \sim \left(\frac{c^{2}}{4 f_\odot}\right) \left(\frac{W}{a}\right) \rho \tau R,
\label{eq_PR}
\end{equation}
where 
$c$ is the velocity of light, 
$f_\odot$ is the solar flux density at Chariklo,
$a$ is the ring semi-major axis,  
$\rho$ the density of the particles and
$\tau$ is the optical depth.
Assuming icy particles, we obtain a typical value of $t_{PR} \sim 10^9 R$~years for
$R$ expressed in metres, i.e. a few million years for sub-cm particles.
The effect of PR drag is even more drastic for smaller grains,
with the fall on the central body in a few million years for 100~$\mu$m-sized particles,
and depletion of micrometric particles by the radiation pressure\cite{Burns79} on a time-scale of months only. 
Even if very crude, this estimation shows that 
Chariklo's rings are either very young, or confined by an active mechanism.

A classical theory invokes the presence of ``shepherd satellites".
In its simplest version, the shepherds confine the edges of the rings
through torques $T_m$ associated with discrete $m+1:m$ mean motion resonances\cit{Goldreich82}:
\begin{equation}
T_m \sim 8.5m^2 a^4 n^2 \Sigma \left(\frac{M_s}{M_C}\right)^2,
\label{eq_torque_m}
\end{equation}
where $m$ is an integer, $a$ is the semi-major axis and
$M_s$ and $M_C$ are the masses of the satellites and Chariklo, respectively.
As $m$ increases, the resonances overlap and the torque density 
(torque exerted per unit interval of semi-major axis) is
\begin{equation}
\frac{dT}{da} \sim 2.5 a^3 n^2 \Sigma \left(\frac{M_s}{M_C}\right)^2 \left(\frac{a}{x}\right)^4,
\label{eq_torque_over}
\end{equation}
where $x$ is the distance between the satellite and the ring.
Those torques must balance the viscous torque caused by inter-particle collisions:
\begin{equation}
T_\nu = 3\pi n a^2 \nu \Sigma,
\label{eq_torque_nu}
\end{equation}
Making $T_m = T_\nu$ in the case of discrete resonances, the radius $R_s$ of a shepherd 
with density $\rho_s$ is: 
$$
\frac{R_s}{R_C} 
\sim \left(\frac{\rho_C}{\rho_s}\right)^{1/3}  \left(\frac{h}{m a}\right)^{1/3} 
\sim {\sf a~few~km}
$$
Concerning the gap between C1R and C2R, it must be opened in the overlapping resonance
regime (Eq.~\ref{eq_torque_over}), in which case, the radius of the satellite is\cit{Goldreich82}
$$
\frac{R_s}{R_C} 
\sim \left(\frac{\rho_C}{\rho_s}\right)^{1/3}  \left(\frac{h}{a}\right)^{1/3}  \left(\frac{W_{\sf gap}}{a}\right)^{1/2}
\lo 1~{\sf km}
$$
where $W_{\sf gap} \sim$ 8.5~km is the full width of the gap, see ED Table~\ref{tab_ring_danish}.

Complications arise because the viscous
torque~(Eq.~\ref{eq_torque_nu}) can be significantly reduced due to the local reversal of the viscous
angular momentum flux, caused by the satellite itself\cit{Goldreich87}. 
This in turn reduces the masses of the shepherds estimated above.
More elaborate theories are beyond the the scope of this work, especially because 
we know nothing about the sizes and locations of the putative shepherd satellites.
The important point, however, is to note that small, kilometre-sized satellites may
confine Chariklo's rings according to the standard model of confinement.


\section{\sf Chariklo size and shape \label{CharikloShape}}

Chariklo's limb is determined by using the ingress and egress timings as given in ED Table~\ref{tab_Chariklo_timings},
and proceeding as for the elliptical fit to the ring events (ED Table~\ref{tab_ring_geo}).
As explained earlier, the SOAR timings may have suffered from  shutter delays and are not used in the limb fitting.
This leaves us with three occultation chords,  two of them (Danish and TRAPPIST) being virtually identical, 
finally providing only two independent occultation chords, one at La Silla and one at Cerro Tololo.



If Chariklo's rings have circular orbits, their projections in the plane of the sky are ellipses 
whose centre should coincide with Chariklo's centre of mass.
If Chariklo is a homogeneous spheroid or ellipsoid, this centre of mass coincides with 
the limb centre.
Forcing the limb and the ring centres to coincide, we are left with 
M~=~3 free parameters for the limb fit, 
its semi-major axis $R_{eq}$, apparent oblateness $\epsilon= (R_{eq}-b)/R_{eq}$ ($b$ is the semi-minor axis) and position angle, to adjust in
N~=~4 independent data points (the four extremities of the occultation chords, see Fig.~\ref{fig_rings}).
The fit is thus overdetermined, and, in general, a satisfactory solution is not expected.

However, we note that the ring centre ($f_{\sf c},g_{\sf c}$) (see ED Table~\ref{tab_ring_geo})
is on the line that connects the centres of the two chords used for the limb fitting.
This means that a solution for which the centre of the elliptical limb coincides
with the centre of the rings does exist.
This is encouraging evidence that the rings are indeed concentric with Chariklo.

We have explored values of 
the apparent oblateness $\epsilon$ from 0 to 0.45 in steps of  0.0025, and 
position angle $P$ from -180\dg\ to 180$^\circ$ in steps of 0.25$^\circ$ (with the centre fixed to $f_{c},g_{c}$).
At each step, the best $R_{eq}$, the corresponding $\chi^2$ 
and the radial residuals are stored. 
From the best solution (reached at the minimum value $\chi^2_{\sf min}$), we obtain 1$\sigma$-error bars
by varying the $\chi^2$ function from $\chi^2_{\sf min}$ to $\chi^2_{\sf min} + 1$.
The best fit corresponds to a projected ellipse with equivalent radius 
$R_{\sf equiv}= \sqrt{R_{eq}\cdot b}= 128.57\pm0.03$~km, 
apparent flattening $\epsilon=0.212\pm0.002$ and 
position angle $P=-53.5\dg \pm0.25\dg$. Note that those are formal 1$\sigma$ error bars from the fit, that may not reflect the real size determination error, as our limb profile result can be affected by topographic features of the order of some kilometres\cit{Johnson73}. The fit also provides the astrometric position of Chariklo with respect to the star (ED Table~\ref{tableChar}).

The best elliptical fit presents very small radial residuals at all the four points (0.7~km), which indicate a satisfactory solution. 
Nevertheless, we see that with $P=-53.5\dg$, the limb is not aligned with the rings, having a mismatch of about 8$^\circ$ (ED Table~\ref{tab_ring_geo}).

If Chariklo is an oblate spheroid, this discrepancy is not expected. 
So we investigate the simplest possible solution, 
where both the centre and the position angle of Chariklo are coincident with those from the rings
 (ED Table~\ref{tab_ring_geo}). 
In this case, we are left with M~=~2 free parameters for the limb ($R_{eq}$ and $\epsilon$)
to fit on N~=~4 chord extremities. 
By the same procedure described above, 
we obtain an elliptical fit with semi-major axis of $\sf R_{eq}=142.9~\pm~0.2$~km 
and flattening $\epsilon=0.213~\pm~0.002$ 
(1$\sigma$ internal error bars),  see ED Table~\ref{tableChar}. 
If Chariklo is an oblate spheroid observed with a tilt angle {\sf B~=~33.77}~degrees (i.e., equal to the ring opening angle), 
then the corresponding true flattening would be 0.330.

The solution's radial residual is almost 4~km (rms), which is fully compatible with the presence 
of putative topographic features in Chariklo's limb\cit{Johnson73}. 
We stress that, due to possible topographic features, 
our effective error bars are of the order of some kilometres.
This fit has an equivalent radius of $R_{\sf equiv}=127$~km 
(and corresponding geometric albedo of 0.031), 
compatible with the thermal measurements\cit{Fornasier13} 
made with the \emph{Herschel Space Telescope}  ($R_{\sf equiv}=124\pm9$~km). Note, however, that Herschel's results are model dependent, and (like our results) may give different values to Chariklo's equivalent radius and albedo if it is an irregular body.

\section{\sf Chariklo Roche limit\label{roche}}

A ring is expected to lie inside its Roche zone, the region where a body 
with critical density $\rho_{\sf Roche}$ is pulled apart by tidal forces\cit{Tiscareno13b}. Note that some minor mixing between small satellites and ring, like the Saturn satellites Pan and Daphnis, is allowed at the boundary. 
The limit $a_{\sf Roche}$ is related (Eq. \ref{eqroche}) to the central object density $\rho_C$ and its equatorial radius $R_{\sf eq}$ (i.e., the limb semi-major axis).

\begin{equation}
\rho_C= \frac{\gamma\ \rho_{\sf Roche}}{4 \pi }\left( \frac{a_{\sf Roche}}{R_{\sf eq}}\right)^3
\label{eqroche}
\end{equation}

The dimensionless parameter $\gamma$ describes the geometry of the assembling components, for instance we have
$\gamma=4\pi/3$ if the particles are spherical; 
$\gamma\sim1.6$ if the material of the body is uniformly spread into its lemon-shaped Roche lobe\cit{Porco07},
and the unlikely low value 
$\gamma\sim0.85$ if we have an incompressible fluid (providing the classical Roche limit).

None of the parameters $\rho_{\sf Roche}$, $\rho_C$ and $\gamma$ are known  (only to factors of several) for the Chariklo system.
Assuming that the rings lie near the Roche limit at $a_{\sf Roche}=400$~km, 
taking $R_{\sf eq} \sim$~145~km from ED Table~\ref{tableChar}, 
we have that $\rho_C=1.67\ \gamma\ \rho_{\sf Roche}$. 
Assuming plausible values for $\gamma$ and $\rho_{\sf Roche}$ 
(knowing that typical values\cit{Tiscareno13} for $\rho_{\sf Roche}$ are 0.5~g~cm$^{-3}$, 
for Saturn's outer ring, and 1.2~g~cm$^{-3}$ 
for Uranus outer rings), 
we see that an ice body density for Chariklo ($\sim$~1~g~cm$^{-3}$) can explain the ring position, 
respecting the Roche limit constraint.


\section{\sf Ring disruption by a planet encounter}


As Chariklo is on an unstable orbit with a short life-time (10 Myr) controlled by Uranus\cit{Horner04}, 
it is instructive
to calculate the encounter distance with the giant planet,  $\Delta_{\sf disrupt}$, 
that would disrupt the ring.
Binary system disruption by an encounter with a giant planet had been studied as a mechanism of satellite capture\cit{AgnorHamilton06}, the closest distance to avoid disruption is given by the classical Hill sphere for a three body encounter:

\begin{equation}
\Delta_{\sf disrupt} \sim a \left(\frac{3M_U}{M_C}\right)^{1/3}
\end{equation}

Here $M_U \sim 10^{26}$~kg is Uranus' mass, $a \sim 400$~km is the ring radius and 
$M_C \sim 10^{19}$~kg is Chariklo's mass, providing 
$\Delta_{\sf disrupt} \sim 0.0008~{\sf AU} \sim$ 5~Uranus radii. More sophisticated analysis, like the effect of the ring orientation with respect to Uranus during the encounter, may change this number but will keep the same order of magnitude.

Simulations that were aimed at studying satellite capture by the giant planets\cit{Nogueira11}, 
had shown that less than 0.001\% of the test objects had close encounters with the giant planets of that order.  

\clearpage

\setlength{\parskip}{0mm}
\setlength{\parindent}{-6mm}



\vspace{-1.5cm}
\begin{thebibliography}{99}
 
\bibitem{Tiscareno13}
Tiscareno, M.~S.\  
Planetary Rings,
in \textit{Planets, Stars and Stellar Systems.~Volume 3: Solar and Stellar Planetary Systems.}
(eds {Oswalt}, T.~D., {French}, L.~M. \& {Kalas}, P.),
309--375 (Springer Netherlands, 2013).

\bibitem{Fornasier13}
Fornasier, S., \etal \ 
TNOs are Cool: A survey of the trans-Neptunian region VIII. Combined Herschel PACS and SPIRE observations of nine bright targets at 70-500 $\mu$m.
\aap, {\bf555}, A15 (2013). 

\bibitem{Belskaya10} 
Belskaya, I.~N., \etal \ 
Polarimetry of Centaurs (2060) Chiron, (5145) Pholus and (10199) Chariklo. 
\icarus, {\bf210}, 472--479 (2010).

\bibitem{Guilbert09}
Guilbert, A., \etal \ 
A portrait of Centaur 10199 Chariklo.  
\aap, {\bf51}, 777--784 (2009).

\bibitem{Guilbert11}
Guilbert-Lepoutre, A. \ 
A Thermal Evolution Model of Centaur 10199 Chariklo. 
\aj, {\bf141}, 103 (2011).

\bibitem{Horner04}
Horner, J.,  Evans, N.~W.  \& Bailey, M.~E. \ 
Simulations of the population of Centaurs - I. The bulk statistics. 
\mnras, {\bf354}, 798--784 (2004).

\bibitem{Camargo13}
Camargo, J.I.B., \etal \ 
Candidate stellar occultations by Centaurs and trans-Neptunian objects up to 2014.
\aap, {\bf 561}, A37 (2014). 

\bibitem{Harpsoe12} 
Harps{\o}e, K.~B.~W., J{\o}rgensen, U.~G., Andersen, M.~I., \& Grundahl, F.\ 
High frame rate imaging based photometry. Photometric reduction of data from electron-multiplying charge coupled devices (EMCCDs).
\aap, \textbf{542}, A23, (2012). 

\bibitem{Skottfelt13}
Skottfelt, J., \etal\ 
EMCCD photometry reveals two new variable stars in the crowded central region of the globular cluster NGC 6981. 
\aap, \textbf{553}, A111, (2013).


\bibitem{French91}
French, R.~G., Nicholson, P.~D., Porco, C.~C. \& Marouf, E.~A. \ 
Dynamics and structure of the Uranian rings, 
in \textit{Uranus.} 
(eds Bergstralh, J.~T., Miner, E.~D. \& Matthews, M.~S.),
327--409 (Univ. Arizona Press, Tucson 1991).

\bibitem{Colwell09}
Colwell, J.~E., \etal\ 
The Structure of Saturn's Rings, 
in \textit{Saturn from Cassini-Huygens.} 
(eds Dougherty, M.~K., Esposito, L.~W. \& Krimigis, S.~M.),
375--412 (Springer, 2009).

\bibitem{Esposito91} 
Esposito, L.~W.,  Brahic, A., Burns, J.~A., \& Marouf, E.~A.\ 
Particle properties and processes in Uranus' rings,
in \textit{Uranus.} 
(eds Bergstralh, J.~T., Miner, E.~D. \& Matthews, M.~S.),
410--465 (Univ. Arizona Press, Tucson 1991).

\bibitem{Karkoschka01}
Karkoschka, E.\ 
Comprehensive Photometry of the Rings and 16 Satellites of Uranus with the Hubble Space Telescope. 
\icarus, {\bf151}, 51--68 (2001).

\bibitem{Hedman13}
Hedman, M.~M., \etal\ 
Connections between spectra and structure in Saturn's main rings based on Cassini VIMS data.
\icarus, {\bf 223}, 105--130 (2013).



\bibitem{Tiscareno13b} 
Tiscareno, M.~S., Hedman, M.~M., Burns, J.~A. \& Castillo-Rogez, J.\ 
Compositions and Origins of Outer Planet Systems: Insights from the Roche Critical Density.
\apjl, {\bf765}, L28 (2013).

\bibitem{Goldreich79}
Goldreich, P. \& Tremaine, S.\  
Towards a theory for the Uranian rings.
\nat, {\bf277}, 97--99 (1979).

\bibitem{Elliot10}
Elliot, J. L.,  \etal \
Size and albedo of Kuiper belt object 55636 from a stellar occultation.
\nat, {\bf465}, 897--900 (2010).


\bibitem{Sicardy11} 
Sicardy, B., \etal\ 
A Pluto-like radius and a high albedo for the dwarf planet Eris from an occultation.
\nat, {\bf478}, 493--496 (2011).

\bibitem{Ortiz12}
Ortiz, J.~L., \etal\ 
Albedo and atmospheric constraints of dwarf planet Makemake from a stellar occultation.
\nat, {\bf491}, 566--569 (2012).

\bibitem{BragaRibas13}
Braga-Ribas, F., \etal\   
The Size, Shape, Albedo, Density, and Atmospheric Limit of Transneptunian Object (50000) 
Quaoar from Multi-chord Stellar Occultations.
\apj, {\bf773}, 26 (2013).

\bibitem{Bus96}
Bus, S.~J., \etal\ 
Stellar Occultation by 2060 Chiron.
\icarus, {\bf123}, 478--490 (1996).

\bibitem{Elliot95}
Elliot, J. L.,  \etal\ 
Jet-like features near the nucleus of Chiron.
\nat, {\bf373}, 46--49 (1995).

\bibitem{Brunini14} 
Brunini, A.\ 
On the dynamical evolution and end states of binary centaurs.
\mnras, {\bf437}, 2297--2302 (2014).

\bibitem{Noll08}
Noll, K.~S., Grundy, W.~M., Chiang, E.~I.,  Margot, J.-L. \& Kern, S.~D. \ 
Binaries in the Kuiper Belt,
in \textit{The Solar System beyond Neptune.} 
(eds Barucci, M.A., Boehnhardt, H., Cruikshank, D.P. \& Morbidelli, A.),
345--363 (Univ. Arizona Press, Tucson 2008).

\bibitem{Ortiz12b}
Ortiz, J.~L., \etal\ 
Rotational fission of trans-Neptunian objects: the case of Haumea.
\mnras, {\bf 419}, 2315--2324 (2012)

\bibitem{Cuk13}
\'Cuk, M., Ragozzine, D.  \& Nesvorn\'y D.\ 
On the dynamics and origin of Haumea's moons. 
\aj, {\bf146}, 89 (2013).

\bibitem{Nogueira11}
Nogueira, E., Brasser, R. \& Gomes, R.\ 
Reassessing the origin of Triton.
\icarus, {\bf 214}, 113--130 (2011).



\clearpage  

\vspace{-1.5mm}
\textbf{References}
\setcounter{page}{11}   


\bibitem{Zacharias13}
Zacharias, N., \etal\ 
The Fourth US Naval Observatory CCD Astrograph Catalog (UCAC4).
\aj,  {\bf145}, 44 (2013).

\bibitem{Jehin11}
 Jehin, E., \etal\ 
TRAPPIST: TRAnsiting Planets and PlanetesImals Small Telescope.
\emph{The Messenger},  {\bf 145}, 2--6 (2011).


\bibitem{Roques87}
Roques, F., Moncuquet, M. \& Sicardy, B. \ 
Stellar occultations by small bodies: diffraction effects. 
\aj, {\bf93}, 1549--1558 (1987).

\bibitem{Cuzzi85}
Cuzzi, J.~N. \ 
Rings of Uranus: Not So Thick, Not So Black.
\aj, {\bf63}, 312--316 (1985).

\bibitem{Cuzzi09}
Cuzzi, J., \etal\ 
Ring particle composition and size distribution,
in \textit{Saturn from Cassini-Huygens.} 
(eds Dougherty, M.~K., Esposito, L.~W. \& Krimigis, S.~M.),
459--509 (Springer, 2009).

\bibitem{Skrutskie06}
Skrutskie, M. F., \etal\ 
The Two Micron All Sky Survey (2MASS).
\aj, {\bf131}, 1163--1183 (2006).

\bibitem{Cutri12}
Cutri, R. M., \etal\ 
WISE All-Sky Data Release (2012+).
\textit{VizieR Online Data Catalog}, {\bf 2311}, 0 (2012).

\bibitem{CastelliKurucz03}
Castelli, F. \& Kurucz, R. L.
New Grids of ATLAS9 Model Atmospheres,
in \textit{Modelling of Stellar Atmospheres.} 
(eds N. E. Piskunov, W. W. Weiss, \& D. F. Gray), 
{\bf ASP-S210}, A20 (IAU Symp., Uppsala, Sweden 2003).

\bibitem{Fitzpatrick99}
Fitzpatrick, E. L. \ 
Correcting for the Effects of Interstellar Extinction.
\pasp, {\bf 111}, 63--75 (1999).




\bibitem{Colwell92}
Colwell, J.~E. \& Esposito, L.~W. \ 
Origins of the rings of Uranus and Neptune. I - Statistics of satellite disruptions.
\jgr, \textbf{97}, 10227--10241 (1992).

\bibitem{Goldreich82}
Goldreich, P. \& Tremaine, S. \ 
The Dynamics of Planetary Rings.
\araa, \textbf{20}, 249--283 (1982).


\bibitem{Burns79} 
Burns, J.~A., Lamy, P.~L. \& Soter, S.\ 
Radiation forces on small particles in the solar system.
\icarus, {\bf40}, 1--48 (1979).



\bibitem{Goldreich87}
Goldreich, P. \& Porco, C.~C. \ 
Shepherding of the Uranian rings. II. Dynamics.
\aj, \textbf{93}, 730--737 (1987).

\bibitem{Johnson73}
Johnson, T.~V. \& McGetchin, T.~R. \ 
Topography on satellites surfaces and the shape of asteroids. 
\icarus, {\bf18}, 612--620 (1973).

\bibitem{Porco07}
Porco, C.~C.,  Thomas, P.~C., Weiss, J.~W. \& Richardson, D.C.\ 
Saturn's Small Inner Satellites: Clues to Their Origins.
\sci, {\bf318}, 1602--1607 (2007).

\bibitem{AgnorHamilton06} 
Agnor, C.~B. \& Hamilton, D.~P.\ 
Neptune's capture of its moon Triton in a binary-planet gravitational encounter.
\nat, {\bf441}, 192--194 (2006).





\end{thebibliography}
\end{document}